\documentclass{article}

\usepackage{microtype}
\usepackage{graphicx}
\usepackage{subfigure}
\usepackage{booktabs} % for professional tables

\usepackage{hyperref}
\usepackage{enumitem}

\usepackage[accepted]{icml2021}
\setcitestyle{square}
\setcitestyle{numbers} %hay que meter esto para que compile bien en arXive por un tema de la bibliografía

\icmltitlerunning{Advances in Agentic AI: Back to the Future}

\makeatletter
\renewcommand{\printAffiliationsAndNotice}[1]{%
  \begingroup
  \renewcommand{\thefootnote}{\fnsymbol{footnote}}%
  \footnotetext[1]{%
    \textbf{Affiliations:} \\
    Sergio Álvarez-Teleña: Department of Computer Science, University College London, UK \\
    SciTheWorld, Spain\\
    Marta Díez-Fernández: SciTheWorld, Spain \\
    Correspondence to: sergio@scitheworld.com, marta@scitheworld.com
  }%
  \endgroup
}
\makeatother
\begin{document}

\twocolumn[
\icmltitle{Advances in Agentic AI: Back to the Future}
\vspace{0.5em}
\begin{center}
Sergio Álvarez-Teleña\textsuperscript{1,2} \quad Marta Díez-Fernández\textsuperscript{2} \\
\end{center}
\vspace{0.5em}

\vskip 0.3in
]



\printAffiliationsAndNotice{}

\begin{abstract}
In light of the recent convergence between Agentic AI and our field of Algorithmization, this paper seeks to restore conceptual clarity and provide a structured analytical framework for an increasingly fragmented discourse. First, \textit{(a)} it examines the contemporary landscape and proposes precise definitions for the key notions involved, ranging from intelligence to Agentic AI. Second, \textit{(b)} it reviews our prior body of work to contextualize the evolution of methodologies and technological advances developed over the past decade, highlighting their interdependencies and cumulative trajectory. Third, \textit{(c)} by distinguishing Machine and Learning efforts within the field of Machine Learning \textit{(d)}  it introduces the first Machine in Machine Learning (M1) as the underlying platform enabling today’s LLM-based Agentic AI, conceptualized as an extension of B2C information-retrieval user experiences now being repurposed for B2B transformation. Building on this distinction, \textit{(e)} the white paper develops the notion of the second Machine in Machine Learning (M2) as the architectural prerequisite for holistic, production-grade B2B transformation, characterizing it as Strategies-based Agentic AI and grounding its definition in the structural barriers-to-entry that such systems must overcome to be operationally viable. Further, \textit{(f)} it offers conceptual and technical insight into what appears to be the first fully realized implementation of an M2. Finally, drawing on the demonstrated accuracy of the two previous decades of professional and academic experience in developing the foundational architectures of Algorithmization, \textit{(g)} it outlines a forward-looking research and transformation agenda for the coming two decades.\\
\end{abstract}

\section{Introduction}
\label{section: Introduction}

\subsection{Context}
\label{subsection: Context}

\textit{Agentic AI} has become a dominant theme in global technological discourse. By early 2025, it had permeated boardrooms, consultancy frameworks, national digital strategies, and the public imagination. Yet, despite its rapid ascension, \textbf{the term has been used inconsistently and often imprecisely} - frequently \textbf{pegged to Large Language Models (LLMs)} methodologies. This widespread misconception has obscured the real architectural and strategic implications of agentic systems and has led corporations, governments, and academic institutions to anchor their expectations on technologies that were never designed for production-grade transformation.\\

This gap became particularly evident in September 2025, when \cite{Why_Hallucinate} disclosed a crucial insight: the phenomenon commonly referred to as “hallucinations” is not an incidental defect but a \textbf{structural property} of LLMs. The errors produced by these models stem directly from their statistical estimation process rather than from any failure of retrieval. As a result, the many corporations preparing to integrate LLM-based Agentic AI into their operational infrastructures - often through superficial or “vibe-driven” development approaches - were, in effect, poised to embed \textbf{structural noise} into their production systems. This adds a new category of risk atop existing operational fragilities and cybersecurity exposures, making a thorough reassessment of those strategic plans imperative.\\

Importantly, this conclusion was \textbf{not new to our readers}. In our September 2023 publication \cite{Digital Washing}, we argued explicitly that there are “no hallucinations in Large Language Models … but errors by design”, because LLMs do not retrieve information but \textbf{estimate and compose} it. The structural nature of LLM errors was therefore predictable - and was articulated - years before the recent institutional acknowledgment.\\

These issues represent only the surface of a much deeper set of misconceptions and methodological misalignments - misalignments that have been further amplified by several tier-one consulting firms and major technology providers. Why, then, do these persistent misjudgments continue to arise across such influential institutions? While later sections of this paper will dissect what we believe to be the true “elephants in the room”, the core explanation can be summarized succinctly: the current technological landscape remains in a state of \textbf{economic disequilibrium}, driven by a pervasive lack of judgment on both the supply side (technology sellers) and the demand side (corporate buyers). As we will show, the persistence of this disequilibrium is not only understandable but structurally embedded in the incentives and knowledge constraints of the field. \\

Given the \textbf{societal stakes} - ranging from misallocated massive investments to unnecessary large-scale layoffs - the need for conceptual precision and architectural clarity is acute. Our aim is therefore twofold: first, to elucidate the \textbf{origins and drivers} of the current misunderstanding surrounding Agentic AI; and second, to \textbf{clarify the path ahead} so that scientists, executives, policymakers, and practitioners can participate more effectively and responsibly in shaping this technological transition. In doing so, we seek to equip organizations with the judgment and tools required to navigate the shift toward Agentic AI with \textbf{accuracy, responsibility, and efficiency}.\\

\subsection{Contribution Summary}
\label{subsection: Contribution Summary}

This paper argues that the confusion arises from a failure to distinguish between the two underlying components of Machine Learning - \textbf{the Learning (L)} and \textbf{the Machine (M)}. The Learning component, defined by Computational and Traditional Statistics, is fundamentally a \textit{scientific} endeavour whose output, the model \textit{f(·)}, has become a commodity due to the global and open nature of academic innovation. \textbf{Competitive advantage cannot be sustained at the level of the model (the L)}. Instead, it must emerge from the \textbf{Machine}: the architectural and algorithmic infrastructure that governs how models are deployed, orchestrated, combined with heuristics, and adapted to real-world operations. This is the reason why Agentic AI, a new way to create software architecture, has become the cornerstone of right-to-win plans - \textbf{more customized and proprietary than ever towards sustaining the corporate competitive advantage}.\\

To formalize this distinction, we introduce the \textbf{Machine Theory of Agentic AI}, which further differentiates between two machines. \textbf{M1} refers to the merge between science and data engineering required to estimate chip-intensive models (such as those required to calibrate LLMs). \textbf{M2} adds to M1’s capabilities a number of features required to create a federated, modular, algorithmic ecosystem through which organizations can operationalize intelligence across departments, processes, and infrastructures. This is, M2 is notably more complex than M1 (it can create M1s for different purposes) and, we argue, the technology that corporations are ultimately looking for.\\

Within M2 we distinguish two types of approaches: LLM-based and Strategies-based.\\

\textbf{LLM-based M2} refers to the current mainstream in Agentic AI. This is, when M2 is intended to be built upon LLM’s vibe coding so that software is created by non-coders and automatically deployed into production. Note that it is the attempt to create an M2 upon the outcome of an M1 created to calibrate LLMs. Thus, bounded by a \textbf{pivotal strategy} of LLM providers who want to move from B2C (retail facing services) to B2B (corporate facing services). However, as several CTOs have noted, this trajectory raises concerns regarding the extent to which \textbf{marketing narratives}, rather than technical feasibility, are shaping expectations. The structural properties of LLMs—hallucinations, opacity, limited determinism - pose inherent constraints on their ability to support production-grade M2 architectures.\\

\textbf{Strategies-based M2} refers to the original Agentic AI from \cite{Data MAPs}. It conceptualizes M2 as a \textbf{top-down architectural discipline}, grounded in, to the best of our knowledge, the most complex digital business that exists today: \textbf{Algorithmic Trading}. As detailed in \cite{Vertical Integration} and the broader Algorithmization series (see Section \ref{section: Literature Review}, below), this perspective rests on the belief that organizations cannot scale their technological capabilities by progressing upward from the simplest use cases. Instead, \textbf{the correct trajectory is the opposite}: start from the most demanding operational and algorithmic contexts and generalize those capabilities \textit{downward} into all business functions.\\

\begin{figure}[ht]
\vskip 0.1in
\begin{center}
\centerline{\includegraphics[width=\columnwidth]{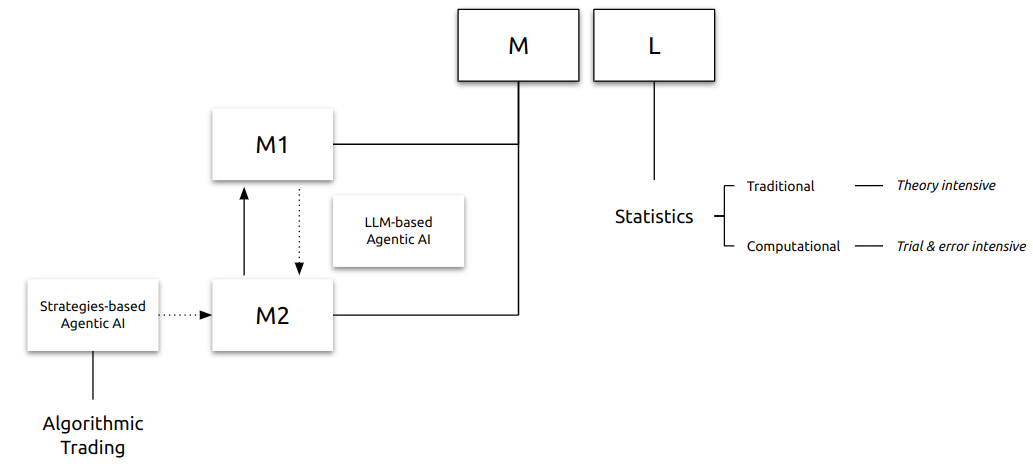}}
\caption{ML's components: an eloquent overview of its nature.}
\label{fig: ML components}
\end{center}
\vskip -0.2in
\end{figure}

The work is inherently interdisciplinary, \textbf{merging Economics, Technology, and Machine Learning} to produce a coherent theory of transformation which can be realistic in terms of business evolution.\\

We conclude by arguing that \textbf{the future of corporate, defense and national transformation will depend} not on the sophistication of statistical models but \textbf{on the mastery of Strategies-based M2}: the ability to combine heuristics, models, data flows, compliance constraints, and federated computational units into a dynamic, resilient, self-orchestrating architecture. In this sense, Agentic AI is not an LLM phenomenon but a more \textbf{complex machine that comprehends a new organizational paradigm} - one that demands clarity, rigour, and deliberate design.\\

\subsection{Scope and Limitations}
\label{subsection: Scope and Limitations}

The scope of this paper is intentionally centered on our past decade evolving \textbf{Applied Science}, a field that remains insufficiently defined precisely because it sits at the intersection of academic rigor and industrial execution. For this reason, even its \textbf{mode of communication} - its vocabulary, structure, and conceptual framing - must diverge from established academic or corporate conventions. Applied Science is intrinsically disruptive: it synthesizes scientific theory with expert heuristics, strategic reasoning with operational deployment, and multiple statistical models with domain-specific judgment to produce solutions that are both impactful and implementable.\\

Crucially, this integration does not encroach upon the domains it draws from. It leaves intact the scientific foundations of \textbf{Science Applied} - the creation of the Ls in ML - along with the engineering and computational challenges embodied in \textbf{M1}, and the domain expertise inherent to \textbf{Applied Science} as instantiated by \textbf{M2}, which comprises business strategy, regulatory constraints, operational design, cybersecurity, and beyond. In doing so, the paper preserves clean conceptual boundaries, enabling each domain to be discussed with appropriate depth and autonomy while still contributing to a coherent overall framework.\\

As a result, the paper occupies a space that is fundamentally atypical: it does not conform to the stylistic or methodological standards of academic research, nor does it follow the conventions of industry whitepapers. It is designed to articulate a discipline that bridges both worlds while being fully contained by neither.\\

Furthermore, this document intentionally incorporates more business insight than any of our prior publications. Note that we do not consider these business-oriented considerations academically elegant.\\

This challenge is also compounded by the proprietary and highly customized nature of such systems, which \textbf{precludes the possibility of standardized quantitative performance comparisons}. As we will explain, such comparisons cannot be meaningfully constructed nor responsibly disclosed - neither by us nor by other actors operating in this niche.\\

However, it is essential that emerging research groups pursuing disruptive innovation - whether in this field or in adjacent domains - gain a clear and unfiltered understanding of the practical realities involved, including both the advantages and the structural challenges inherent in embarking on such a quest. Only with this level of transparency can future researchers calibrate their expectations, assess the risks properly, and prepare themselves for the multidisciplinary demands that define Applied Science at scale.\\

The integration of all this information functions as an experiment in its own right. Given the breadth and depth of the topics involved, we have intentionally structured the material so that readers may benefit from interacting with it through LLMs. When subjects span multiple disciplines and exceed an individual’s areas of expertise, such tools can assist in drawing connections, clarifying dependencies, and navigating the conceptual landscape more effectively. We have observed that when widely used LLMs were asked to critique the document they tended to overlook the fact that \textbf{our claims are grounded in results already described in previous papers} (e.g., size of the companies where we have proved scalability and novel partnerships such as those at institutional level). As detailed in the following chapter, these statements are supported by a broad and diverse body of prior work, and can therefore be eliminated by explicitly uploading the underlying set of papers on which they build. Moreover, when prompted to respond to most criticisms, these systems frequently incorporate the very information that had been omitted or disregarded when building the initial critique, thereby implicitly acknowledging the incompleteness of the original assessment.\\

As will be discussed, the technology described in this work cannot be directly validated by the open-source community, as its development and evaluation require substantial financial, computational, and organizational investment and therefore remain proprietary. Nevertheless, the underlying principles and methodological foundations - articulated here within the framework of \textit{Algorithmization} - are fully reproducible in the scientific sense. Independent research groups can replicate and extend these results by reconstructing comparable systems, in much the same way that Algorithmic Trading emerged and matured through cumulative advances grounded in a shared body of theoretical and empirical literature.\\

The scalability of \textit{Algorithmization} is intrinsically constrained by the extent and intensity of the professional services - namely strategic advisory and data science - delivered by consultants on top of the M2 platform described herein. Consequently, this work deliberately restricts its scope to the M2 layer itself, which constitutes the primary mechanism through which Algorithmization can be scaled globally, independently of human-service bottlenecks.\\

Detailed discussions of the underlying technologies, methodologies, and implementation techniques fall outside this scope. Readers seeking deeper technical insight may refer to our broader \textbf{Algorithmization} corpus, summarized in Section \ref{section: Literature Review}, where these foundations are developed comprehensively.\\

\subsection{Conclusions}
\label{subsection: Conclusions 1}

Meaningful, scalable, and resilient transformation is a Machine problem, not a Learning problem. Agentic AI, correctly understood, is not an extension of LLM dynamics but the architectural discipline required to operationalize intelligence - human and machine -across entire organizations, sectors, and nations. Only through mastery of Strategies-based M2 can institutions transition from superficial digitalisation to genuine competitiveness and structural efficiency.\\

In other words, \textbf{no organization can handle AI effectively until it has first transformed itself}. AI adoption presupposes an operating model, an architecture, and a governance framework capable of integrating it. As we will demonstrate, organizational transformation and AI integration need not proceed sequentially. \textbf{AWhen designed correctly, they can evolve in parallel}, avoiding the loss of momentum that arises from treating them as isolated, successive phases.\\

\section{Literature Review}
\label{section: Literature Review}

Our literature has evolved in parallel with our understanding of \textbf{extreme-efficient transformation}, a paradigm that draws simultaneously on the principles of the \textit{common minimum multiple} and the \textit{fractal form}. At its core lies a fundamental question: how can the largest functional breadth be achieved with the smallest number of architectural components, supported by a rigorous and generalizable methodology?.\\

Accordingly, our roadmap follows a \textbf{gradual and hierarchical progression}, reflecting the natural scaling of transformation itself:
\begin{enumerate}
    \item Products
    \item Departments
    \item Companies
    \item Sectors
    \item Countries
    \item Societies
\end{enumerate}

Our priority was to construct a holistic framework for an \textbf{entirely new discipline} - one that required \textbf{disruptive} rather than progressive innovation. Incremental advances would have been insufficient: the ambition was not to refine existing methodologies but to redefine the architectural, conceptual, and operational foundations upon which transformation should rest. Yet, disruptive innovation is notoriously difficult to fund in both industry and academia. The magnitude of the intellectual and technological leap required a level of commitment that is seldom supported by traditional research grants, corporate budgets, or venture-capital timelines.\\

Further, we recognized that the same barriers-to-entry that constrained us would likewise deter others; the discipline would simply not materialize unless someone was willing to assume the full professional risk required to build it. This is, its pioneers would have to confront \textbf{obstacles extending far beyond the traditional boundaries of research}.\\

A defining tenet of our research quality commitment is that, overall, white papers shall only be published once the underlying technology has first been built and validated in industry. We consider this sequencing essential to prevent expectation misalignment among readers. Furthermore, our long-run ambition required that both our research papers remain \textbf{organically coherent yet maximally orthogonal}. Just as we avoided selling the same product twice in the industry - deliberately seeking new projects to expose ourselves to novel constraints and design challenges - we also avoided incremental academic updates.\\

All above allowed us to earn the trust of numerous Tier-1 institutions and, in turn, enabled us to bootstrap our entire Centre of Excellence (CoE) - we believe this crucially constitutes another affirmation of the credibility and relevance of our work in \textbf{Applied Science}, where the notion of \textit{peer review} extends across a broad and heterogeneous set of academic and market participants.\\

It is upon the whole literature framework of Applied Science, \textbf{Algorithmization}, that we expect that \textbf{incremental innovation} will emerge naturally in the future, particularly through collaborations with academic partners.\\

\subsection{Products}
\label{subsection: Products}

The products we created were designed through a \textbf{SaaS-based framework}. Because our objective was to develop an AI-first, all-in-one corporate software platform - flexible, federated, and continuously improvable both within and across organizations - we were required to leverage \textbf{Custom SaaS} nearly nine years before the concept entered mainstream discourse and was later popularized under the broader label of \textit{Software-as-a-Service} (still an industry challenge).\\

\subsubsection{Pre-CoE}
\label{subsubsection: Pre-CoE}

Before the creation of the Centre of Excellence, we had prior work that - although not conceived within a Custom SaaS paradigm - proved highly relevant for the technological trajectory that followed.\\

In \cite{PhD}, a set of solutions was introduced that remain advanced even by contemporary standards.\\

It introduces an algorithmic methodology that challenges the traditional microeconomic principle asserting that, under perfect competition, equilibrium prices converge to marginal costs. By focusing on a fully digital-native product - equity index futures - the PhD thesis demonstrates that novel pricing mechanisms can be engineered to achieve deeper (statistical) efficiencies than those implied by classical theory. In doing so, it shows that when markets are sufficiently digitized, new algorithmic structures can shift the attainable efficiency frontier, \textbf{revealing pricing dynamics that were previously inaccessible} within standard microeconomic frameworks - most notably, full replication hedging.\\

The thesis also exposed a divergence between academic developments and business impact. At the time, academic research increasingly favoured elegant combinations of methodologies without sufficiently emphasizing domain-specific expertise. By reframing, with expert insights, the solution to a well known trading problem, the thesis was able to \textbf{outperform the prevailing academic state of the art} modelling with more nimble approaches. This marked the beginning of our distinction between \textbf{Science Applied} and \textbf{Applied Science}, a separation that we develop in detail below.\\

Last, it introduced \textit{Avatar Calibration}, which, to the best of our knowledge, constituted the first application of an \textbf{Augmented Machine} framework in the field - a theme to which we return later.\\

It is important to note that an early book version of \cite{PhD} was released in 2012 (see \cite{book}). Because the doctoral work had been completed ahead of schedule and had to wait for years before defending it, there was a desire to make its results available to industry without delay. The book ultimately circulated across all Tier-1 U.S. investment banks and several leading hedge funds. Its title, \textit{Trading 2.0: Learning-Adaptive Machines}, already hinted at the growing importance we attributed to the \textbf{Machine} in Machine Learning, foreshadowing the line of inquiry that would later culminate in the Algorithmization programme.\\

Industrially, the thesis catalyzed the creation of a new department at BBVA - \textit{to the best of our knowledge, the first sell-side unit to centralize and algorithmically manage all strategies across all asset classes, globally}. This body of work also underpinned the development of the platform that subsequently received the \textbf{Banking Technology Award (2016)} for Best Trading Platform in Europe, thereby demonstrating the practical relevance and industry-grade robustness of the underlying methodology.\\

\subsubsection{Post-CoE}
\label{subsubsection: Post-CoE}

\textit{Data MAPs}, as articulated in \cite{Data MAPs}, became the cornerstone of the Algorithmization framework. Our intention was to first produce a \textbf{deep, conceptually dense paper} - one that would not necessarily be easy to follow - so that subsequent work could reference it while offering more accessible narratives. The paper proposed a fundamentally new way of creating technology - precisely the \textbf{M2, Strategies-based Agentic AI} approach as described below - and sought to demonstrate its universal applicability.\\

At the time, virtually all external feedback - whether from consultants, BigTech firms, venture capital investors, or corporates (with Academia as the only exception) - expressed skepticism that a single architectural paradigm could serve as the \textbf{backbone of any company}. The prevailing expectation was for \textbf{laser-focused products built upon equally narrow technologies}, mirroring the nature of M1 as also described below. No one had attempted, let alone demonstrated, a more ambitious and generalizable approach.\\

In response, we constructed \textbf{The Cube}, a multidimensional framework populated with concrete examples of products delivered across a wide range of industries, organizational sizes, and levels of digital maturity, each annotated with its respective Technology Readiness Level (TRL). Only once we believed The Cube was \textbf{sufficiently populated to substantiate the universality and scalability of deployment of the architecture} did we release the white paper.\\

The vision was deliberately ambitious: \textbf{off-the-shelf technology inevitably generates over-complicated stacks}, producing organizations with \textbf{weak rights-to-play} rather than \textbf{strong rights-to-win}. We instead envisaged a world in which technology is not merely assembled but \textbf{crafted} - a world where firms transcend generic, commoditized tools and construct \textbf{lean, intelligent, and strategically coherent infrastructures}. Through proprietary architectural design and deliberate differentiation, such firms would strengthen their \textbf{right-to-win}, securing a competitive position unattainable through standardised technological offerings.\\

The results were promising: we identified a method - arguably, a breakthrough - to \textbf{scale technological customization} in a manner that remained fully \textbf{interoperable across all dimensions of The Cube}. This breakthrough opened a wide range of novel—and previously unimaginable - possibilities. Most notably, it enabled the construction of an \textbf{ultra-lean, fully proprietary technological infrastructure} that can be deployed with minimal human overhead while remaining \textbf{competitive with the end-to-end stacks of listed companies} (with hundreds of millions of budget to run their infrastructures). Crucially, this approach repositions the locus of value from rights-to-play - a consequence of the widespread reliance on identical, off-the-shelf software stacks - to rights-to-win, deriving from proprietary architectural design and strategic differentiation, much as observed in \textbf{Algorithmic Trading}.\\

Given its disruptive nature - at odds with the incremental innovation typically favored by indexed academic journals - and given the extensive discussion of SciTheWorld as an operating entity, we anticipated that SSRN validators would interpret the manuscript as self-promotional rather than as the documentation of an industrial-scale experiment\footnote{See Appendix for standard criticisms.}, and therefore reject it - as ultimately occurred. This outcome is consistent with well-documented cognitive and institutional biases in human evaluation processes, including incentive misalignment, novelty aversion, and reputational risk management. But it is also fair in academic transformation just as there is change resistance for a myriad of reasons in industry transformation.\\

Notably, we hypothesize that large language models (LLMs) may, under controlled conditions, be more accommodating to genuinely novel constructs than human evaluators. In any case, we could cut the document down to the Machine Theory of Agentic AI in Section \ref{section: The Machine Theory of Agentic AI: M1 and M2} in order to be compliant or write a book instead - both are valid options that we may consider going forward.\\

\begin{figure}[ht]
\vskip 0.1in
\begin{center}
\centerline{\includegraphics[width=\columnwidth]{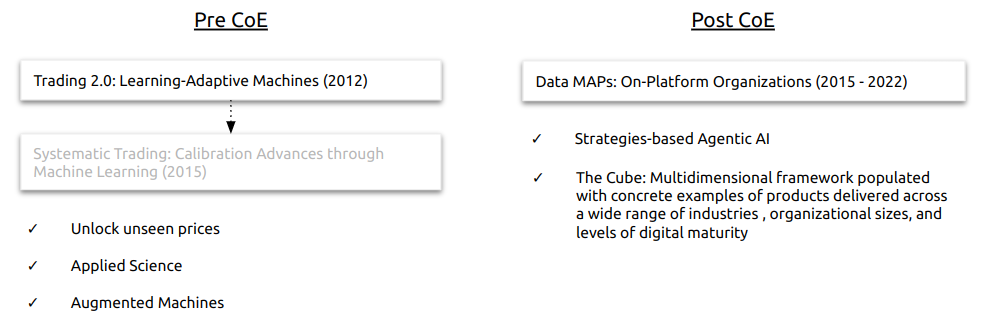}}
\caption{Highlights of foundational contributions: pre-CoE and post-CoE papers advancing product.}
\label{fig: Products pre and post CoE}
\end{center}
\vskip -0.2in
\end{figure}

\subsection{Departments}
\label{subsection: Departments}

Once several products within a department are transformed, it becomes possible to \textbf{reshape the operational logic of the department itself}. Our experience indicates that many business inefficiencies originate from workflows inherited from prior generations of managers - or even academic prescriptions - whose processes were designed to accommodate the limitations of legacy technologies. As those technologies have since become obsolete, many departments now operate below their potential, leaving substantial value untapped.\\

\cite{Data MAPs} documents multiple instances in which departmental transformation emerged naturally once foundational products were rebuilt through algorithmic architectures.\\

More significantly, the Data MAPs paradigm enables \textbf{new cross-departmental workflows} that were previously impossible. Examples include those explored in \cite{OTAN} - notably across communications and CFO functions to mitigate risks of stock-market and social media manipulation - and in \cite{Modern Cybersecurity} - where cybersecurity, business operations, and compliance functions are jointly orchestrated to bootstrap more efficient budgeting and oversight processes.\\

From an industrial perspective, we are currently supporting the transformation of organizations ranging from approximately 50 to over 10,000 employees, primarily through the strategic sequencing of interventions across departments. Typically starting from the areas that account for the most complete set of challenges and then letting other areas inherit and evolve in a federated manner.\\

As will be formalized in Section \ref{section: Insights of the First M2}, this transformation process comprises three distinct categories of services, each characterized by fundamentally different scalability properties.\\

\textbf{First, strategy consultancy} focuses on determining the optimal prioritization and ordering of transformation initiatives across the organization. This activity is inherently non-scalable, as it relies on high-context judgment, domain specificity, and close interaction with senior management. Moreover, from the firm’s perspective, it is neither desirable nor efficient for such strategic guidance to be monopolized by a single provider; diversification of strategic viewpoints is critical to avoid lock-in, cognitive bias, and path dependency. Accordingly, our framework explicitly leaves space for independent strategy consultancies to assume responsibility for this layer.\\

\textbf{Second, data science consultancy} typically operates at the interface between business intent and scientific formalization, translating strategic objectives into applied models, data pipelines, and experimental validation. Like strategy consultancy, this activity does not scale structurally, as it depends on bespoke problem formulation, iterative exploration, and specialized expertise. Here again, diversification across providers is beneficial for the client, both to mitigate model risk and to avoid methodological monoculture. We therefore intentionally position this function outside the scope of the scalable core.\\

\textbf{Third, custom SaaS delivery} consolidates the outputs of the preceding layers into production-grade systems, embedding Applied Science into business-as-usual operations. Unlike the previous services, this layer is \textbf{intrinsically scalable}: once the underlying architecture, protocols, and governance mechanisms are established, they can be deployed, maintained, and upgraded across organizations with limited marginal cost. This role is analogous to the historical scaling of spreadsheet software - such as Excel - which enabled both employees and consultants to operationalize complex logic through formulas and scripting (e.g., Visual Basic). Thus, we assume responsibility for this scalable layer, including its long-term maintenance, upgrades, access control, and permission strategies.

\subsection{Companies}
\label{subsection: Companies}

Once a sufficiently large number of departments has been transformed, the organization as a whole can become fundamentally different. The cumulative effect of departmental algorithmization does not merely improve isolated workflows; it reconfigures the operational fabric of the firm, enabling forms of coordination, efficiency, and strategic coherence that were previously unattainable.\\

In the early stages of this transition, many companies expressed \textbf{optimism regarding their ability} to develop the necessary technology and algorithmic capabilities internally. A similar pattern had occurred earlier in the trading industry, where quantitative researchers initially attempted to build end-to-end solutions themselves. More recently, data scientists adopted a comparable stance in corporate environments. The subsequent dynamics - senior management eventually deciding to \textbf{discontinue internal efforts and instead recruit external technology providers or highly specialized talent} - mirror the trajectory observed in financial markets two decades earlier. Paper \cite{Digital Washing}, show in detail part of these dynamics in the context of a risk-reward framework.\\

\begin{figure}[ht]
\vskip 0.1in
\begin{center}
\centerline{\includegraphics[width=\columnwidth]{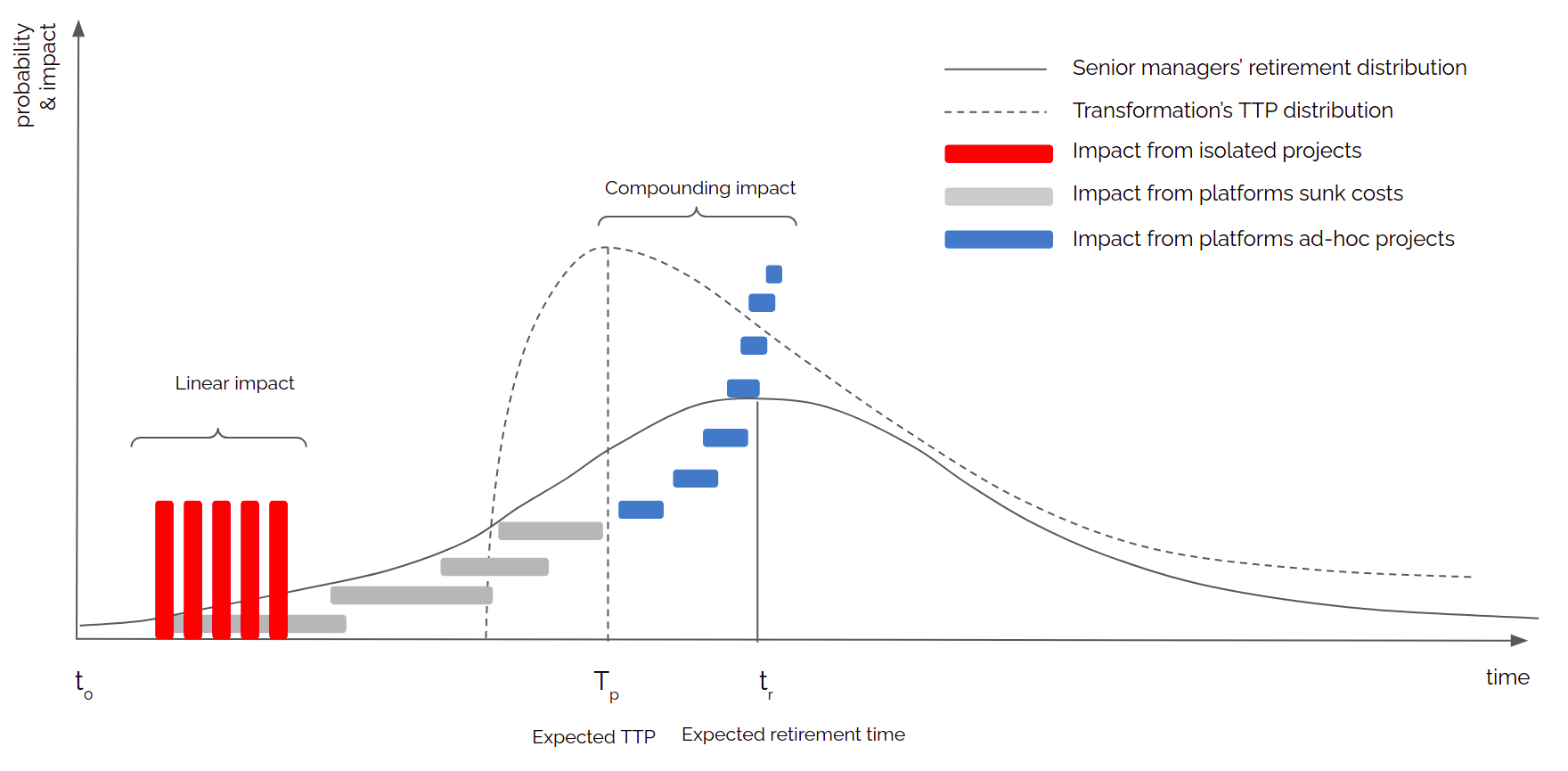}}
\caption{Senior management decision making framework in \cite{Digital Washing}.}
\label{fig: paper CEO}
\end{center}
\vskip -0.2in
\end{figure}

However, as experience across industries demonstrates, transformation at this depth is extraordinarily complex, and the decisive challenges lie in the details. Only a very small number of individuals or teams possess the interdisciplinary mastery required to architect and execute such transformations successfully. This reality aligns with the concentration of capability observed in the hedge fund industry and, more recently, with the emergence of \textbf{above-billion-dollar compensation packages} in firms such as Meta - further evidence of the scarcity and strategic value of elite algorithmic and architectural talent. As we will show, these types of offerings correspond to the \textbf{project management of the architectural logic of M1} within the Machine Learning framework.\\

Once talent and providers are properly selected, the frameworks developed in \cite{3layer}, \cite{OTAN}, and \cite{Revista Infra} offer a \textbf{structured managerial foundation for taking control of the Algorithmization process} and, ultimately, for guiding the organization’s evolution into a fully on-platform enterprise. We developed these frameworks to manage the Centre of Excellence itself, ensuring that its operations remained anchored in principles of control, compliance, and architectural coherence.\\

These works introduce the \textbf{Three-Layer Company} model, which distinguishes among:
\begin{enumerate}
    \item \textbf{M2} (the architectural layer enabling federated, Strategies-based Agentic AI),
    \item the transformation required to preserve the \textbf{right-to-play}, and
    \item the transformation required to protect - and expand - the \textbf{right-to-win}.
\end{enumerate}

The framework further demonstrates its relevance in the context of private equity portfolio management. Its contribution extends beyond the transformation of individual companies: it shows how native interconnectivity across portfolio firms can create \textbf{ecosystem-level synergies}, enabling new strategic narratives for private equity. This is particularly relevant in sectors such as defense, where coordinated algorithmic infrastructures can unlock substantial cross-company value. We will be devoting significant research efforts to this endeavour going forward.\\

Consistent with the notion of integrating companies into a unified ecosystem, the most extreme form of this idea - \textbf{full vertical integration} - is explored in \cite{Vertical Integration}. However, the framework developed therein departs significantly from traditional microeconomic theory. \textbf{Rather than defining verticality through input–output relationships, the paper conceptualizes it through core technological hierarchy}. Under this view, the business with the highest technological complexity becomes the natural integrator of all others - an approach we termed \textbf{Top-Down Vertical Integration}.\\

This perspective also illuminates why, in our view, many \textbf{neo digital banks remain far from efficient} and are not substantively differentiated from incumbent institutions. Their distinguishing feature often lies merely in the use of modern programming languages rather than COBOL - an undeniable burden in legacy finance - but not in possessing a fundamentally superior technological architecture capable of shifting their right-to-win.\\

Finally, in \cite{ASI}, we examine the topic of \textbf{Artificial General Intelligence (AGI)}, a concept that has recently been significantly distorted by commercial narratives. We argue that humanity remains far from engineering any system that resembles \textbf{human-level intelligence} - particularly given our operational definition of intelligence in \ref{subsection: Terminology: A Contextual Framework for Conceptual Clarity} and the still extraordinary energy efficiency of the human brain.\\

\begin{figure}[ht]
\vskip 0.1in
\begin{center}
\centerline{\includegraphics[width=\columnwidth]{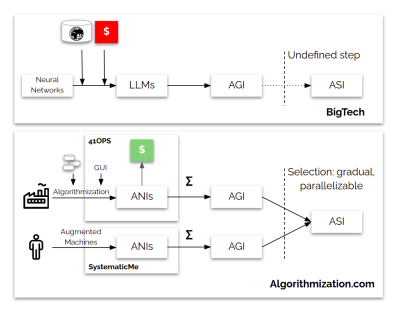}}
\caption{Schematic representation of the rationale behind AGI upon LLMs vs enterprise Algorithmization and Augmented Machines is presented in \cite{ASI}.}
\label{fig: Blockchain vs Data MAPs}
\end{center}
\vskip -0.2in
\end{figure}

However, we propose a fundamentally different perspective by distinguishing between \textbf{Human AGI} and \textbf{Corporate AGI (CAGI)}. The latter is \textbf{synthetic by design}, unconstrained by biological analogy, and its early-stage foundations can already be articulated. CAGI constitutes a genuine \textbf{greenfield for innovation}, capable of enabling substantial global efficiency gains precisely because its architecture allows for \textbf{native interconnectivity} across systems, firms, and industries - an idea first anticipated in \cite{Data MAPs}, especially in contrast to blockchain architectures.\\

\begin{figure}[ht]
\vskip 0.1in
\begin{center}
\centerline{\includegraphics[width=\columnwidth]{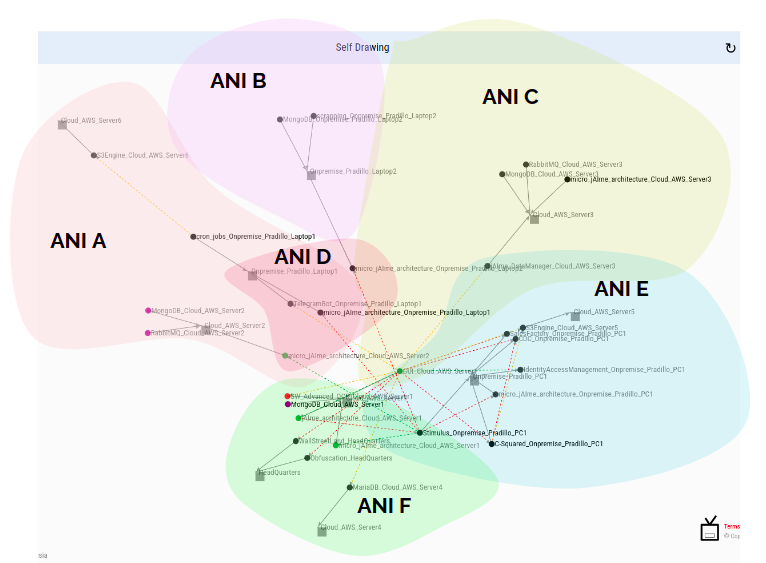}}
\caption{Schematic representation of an Agentic AI platform in areas of Artificial Narrow Intelligences as introduced in \cite{ASI}.}
\label{fig: Brain ANIs}
\end{center}
\vskip -0.2in
\end{figure}

\begin{figure}[ht]
\vskip 0.1in
\begin{center}
\centerline{\includegraphics[width=\columnwidth]{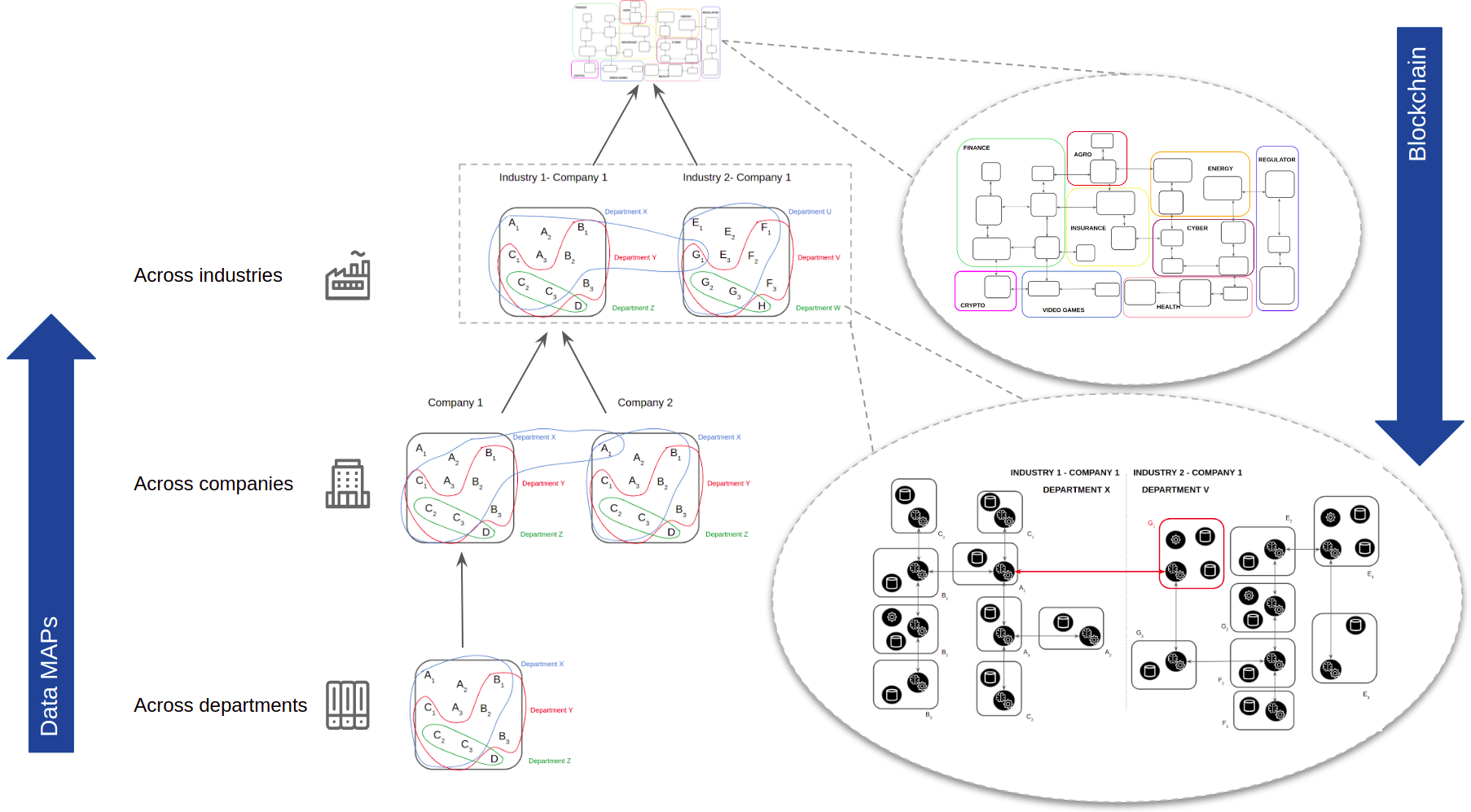}}
\caption{Comparison Data MAPs with Blockchain from \cite{Data MAPs}.}
\label{fig: Blockchain vs Data MAPs}
\end{center}
\vskip -0.2in
\end{figure}

\textbf{Portfolios of companies, as well as large corporate groups} composed of numerous subsidiaries with duplicated departmental structures and duplicated technologies stand to benefit disproportionately from their own CAGI’s emergence. Crucially, these gains \textbf{do not require the centralization} of strategic decision-making across units. Instead, they arise from allowing each entity to retain autonomy while operating within a \textbf{federated innovation architecture}, enabling coordination without imposing uniformity.\\

\subsection{Sectors}
\label{subsection: Sectors}

When you transform several companies from the same industry you change the sector.\\

It is basically the same mindset used above. However, that’s not an easy target to achieve in practice by our CoE within a decade. Hence, instead of proving our theory with technology and success cases, we had to use theory upon our tested technology - the use cases were bound to only our own usage, not with a significant number of companies that allowed us to universalize the impact across the sector.\\

In particular, we proposed modern theory for a couple of sectors - investments and cybersecurity - as a new perspective for researchers to build upon - surely, to be combined with previous approaches, optimally.\\

\subsubsection{Investments}
\label{subsubsection: Investments}
While deploying M2 within a large asset manager, we systematically challenged several long-standing methodological conventions. Many of these practices - though widespread - proved to be inefficient or conceptually flawed, largely because they had been shaped by \textbf{technological limitations of the past}. This motivated the development of a \textbf{trilogy of papers} on technology-modernized portfolio management. Our work focused on three main areas.\\

In \cite{DDP}, we argued that the widespread practice of aggregating positions within a portfolio for the purpose of risk management is not an optimal design choice, but rather a legacy constraint imposed by earlier generations of technology. Historically, because investors lacked the ability to manage each of the underlying strategies that produced individual trades, they were forced to rely on aggregate KPIs at the instrument or portfolio level. This implicitly assumed that financial instruments were sufficiently fungible to justify aggregation.\\

We argue that such fungibility holds only when instruments are similar across a far broader set of dimensions than their basic financial nature - specifically, they must share the strategic rationale that motivated their inclusion (for example, whether one instrument served as a proxy for another). To address this structural limitation, we introduced \textbf{Dimension-Driven Portfolios (DDP): a reframing of portfolio construction and risk control around strategy dimensions, rather than instruments}. This approach proposes an algorithmic-native architecture for portfolio design, enabling more precise, granular, and interpretable management of risk and performance.\\

In \cite{Performance Attribution}, we introduce a new framework for portfolio management that enables investors to interpret and control their exposures through:
\begin{enumerate}
    \item the traditional \textbf{reference benchmark (Beta)};
    \item an \textbf{enhanced version of the benchmark (Alpha-1)}, designed to integrate seamlessly into the statistical-arbitrage ecosystem so that the full body of that theory can be leveraged by design; and
    \item a set of \textbf{free strategies}, encompassing any proprietary trading activity up to and including algorithmic market making \textbf{(Alpha-2)}.
\end{enumerate}

Under this framework, asset managers, investment banks, and hedge funds differentiate themselves not through incompatible methodologies but through the \textbf{allocation of their risk appetite across these three components}. The same underlying technology and methodological architecture apply across all types of institutions. This common structure improves not only an asset manager’s ability to control aggregate risk and clarifies performance attribution, but \textbf{it also enhances other key dimensions of the firm} such as its capacity to attract and retain talent - particularly individuals who are more oriented toward algorithmics than toward traditional finance.\\

In \cite{The Governance}, we extend this line of work by showing how organizations can \textbf{leverage the expertise of portfolio managers while freeing them from operational burdens}, allowing them to devote more time to research and strategic development, while working more coordinatedly. This approach operationalizes the \textbf{Augmented Machines} concept we introduced in 2012 (see \cite{book}): the idea that new roles shall involve human intelligence guiding and supervising machines, while machines autonomously assume scalable and operationally intensive tasks of increasing sophistication.\\

\subsubsection{Cybersecurity}
\label{subsubsection: Cybersecurity}
The initial projects undertaken by the Centre of Excellence were focused on \textbf{advanced cybersecurity}, and this choice proved strategically significant in two distinct ways. First, we intentionally selected a domain far removed from finance to avoid being typecast as a purely financial-technology initiative. Although the co-founders had prominent careers in finance, relying on that legacy would have been the path of least resistance but not the optimal foundation for a discipline intended to transform multiple sectors. Second, beginning with cybersecurity ensured that the technology would be \textbf{architected from inception with maximal cybersecurity flexibility}, forcing us to acquire deep domain knowledge in order to contribute genuinely orthogonal value.\\

As Algorithmization began to gain wider recognition, the Spanish association of CISOs (ISMS) invited us to articulate our perspective on cybersecurity, particularly given the breadth of new capabilities unlocked by Strategies-based Agents. The resulting \cite{Modern Cybersecurity} made several contributions. It identified \textbf{new high-priority risks} that must be jointly considered by business and cybersecurity teams; reframed the resourcing of cybersecurity around the \textbf{hybridization of business, compliance, and security functions}; demonstrated how an algorithmic-native platform enables harmonious orchestration across business and cybersecurity processes; and established a blueprint for future cybersecurity strategy in which \textbf{business continuity and continuity-at-risk} become tightly coupled through tactical technology.\\

\subsection{Countries}
\label{subsection: Countries}
Once sectors undergo transformation at scale, the conditions emerge for transforming a \textbf{nation} as a whole.\\

Nations must increasingly treat technological transformation as a \textbf{geostrategic asset}. As we will discuss, \textbf{M1 is primarily concerned with the creation of AI} - where most contemporary geostrategic competition is currently concentrated. Our work shifts attention toward \textbf{AI usage}, placing emphasis on \textbf{M2}, where industry competitive advantages reside. These advantages lie neither in the statistical models (\textit{L}) nor in the infrastructures that produce them (\textit{M1}), but in the \textbf{architectural capacity to consume AI effectively}. Recent global business experience has confirmed this view: consuming AI - operationalizing it across industries, infrastructures, and institutions - seems to be the most complex and strategically consequential task of all.\\

In examining the national implications, we focused on two key dimensions:
\begin{enumerate}
    \item \textbf{Novel defense}, approached in ways not yet explored in traditional strategic thinking; and
    \item \textbf{GDP enhancement}, through structural productivity gains unlocked by algorithmic-native industry transformation.
\end{enumerate}

\subsubsection{Novel Defense}
\label{subsubsection: Novel Defense}

The Spanish Armed Forces approached us to consider preparing a paper for NATO on advances in AI. Rather than following the well-trodden path that most contributors were exploring, we elected to present a \textbf{fundamentally different idea}: that a country could take over another \textbf{without firing a single shot} - solely through its \textbf{boardrooms}.\\

We demonstrated how, using our technology, a sophisticated attacker could orchestrate \textbf{coordinated market and social-media manipulation}. A hacking-based trading strategy targeting a listed company could execute a highly calibrated attack with two simultaneous outcomes: first, the attacker could profit financially, reinvesting returns into subsequent more ambitious operations; second, by accumulating a sufficiently large position, the attacker could acquire a board seat, thereby \textbf{steering the company’s strategic direction in ways aligned with the interests of the attacking nation}.\\

We further showed how, with technology capable of orchestrating multiple departments within a firm, a \textbf{defensive counter-strategy} could be mounted across social media and financial markets. Importantly, we designed these countermeasures such that they would operate in a manner that \textbf{regulators should find acceptable}, aligning defensive action with compliance expectations in a systematic, transparent, traceable, pre-configurable, auditable context.\\

NATO ultimately did not accept the proposal - a fact that remains concerning, particularly given that the article is now public and thus the strategy can, in principle, be executed by any sufficiently capable adversarial actor.\\

\subsubsection{Extreme-Efficient Nations}
\label{subsubsection: Extreme-Efficient Nations}

By late 2024, as AI increasingly emerged as a central axis of geopolitical competition, we decided to accelerate the publication of a paper originally planned for the late 2020s. At that time, the United States and China were concentrating a significant amount of their strategic efforts on AI creation, competing primarily through advances in model development and computational capacity. It therefore became important to articulate - especially in a moment of heightened geopolitical tension - that a second and equally consequential evolutionary path exists: AI \textit{consumption}, the domain of M2. This paper forms part of that broader effort.\\

The argument was \textbf{twofold}. First, AI consumption represents a major source of economic value, arguably more profitable and scalable than AI creation itself. Second, it is a path that remains strategically aligned with both the United States and China: these countries will ultimately become the primary consumers of the AI technologies the United States and China produce. Moreover, the widely cited statistic that 95\% of AI projects fail (see MIT study\footnote{\href{https://mlq.ai/media/quarterly_decks/v0.1_State_of_AI_in_Business_2025_Report.pdf}{State of AI in Business 2025.}}) illustrates both the limitations of an AI \textit{creation-centric} geostrategy and the global need for sophisticated capabilities in AI consumption. An M2 deployed across countries would benefit the world in an unseen manner\footnote{We introduced the concept of building vs consuming AI in a geostrategic framework at a major think tank meeting (IADG) in June 2025.}.\\

Shortly after publication, we were contacted by the Ministry of Economy of Spain through the Director-General for Economic Policy, to explore the practical implementation of the framework. The Ministry convened key institutional actors, including the Instituto de Empresa Familiar (Spain’s association of family-owned businesses), the Real Instituto Elcano (a leading geopolitical think tank), and the ICO (the State-owned financial agency). This assembly brought together the users of the technology, an agent capable of elevating the initiative from political to institutional status, and an instrument for deploying economic incentives, respectively. \textbf{The first step was to build bottom-up, ad-hoc and timely economic policy through the companies’ process of Algorithmization}.\\

We recognize that \textbf{sustaining such an initiative over time will be challenging}. Changes in personnel across the participating institutions - and even shifts in governing political parties - pose structural risks to continuity. Nonetheless, the strategic relevance of the initiative is such that it must endure. As in \cite{ASI}, maintaining calm, sequencing, and strategic patience will likely be essential. Encouragingly, interest from other countries suggests the potential for a broader international effort toward operationalizing M2 at national scale.\\

\subsection{Societies}
\label{subsection: Societies}
Our first attempt to influence society more broadly is reflected in \cite{book}. When we joined the Computer Science Department at University College London - after having already built professional careers, particularly in algorithmic execution trading - we recognized that a major shift in the relationship between humans and machines was imminent. We also feared that this shift would unfold incorrectly: a pendulum swing from a world of “all people and no machines” to one of “all machines and no people”. We anticipated that this misunderstanding would be fueled by a conceptual confusion between Applied Science and Science Applied, a distinction we elaborate later.\\

It was in this context, back in 2012 (see \cite{book}, that we introduced \textbf{Avatar Calibration}, the most advanced instantiation to date of the \textbf{Augmented Machines} framework to our knowledge. Avatar Calibration demonstrated that a human expert could retain a meaningful role in environments increasingly dominated by machines by \textit{proving to the machine} - through the calibration process - that its performance improved with the human rather than without one. Beyond this contribution, we were dissatisfied with the prevailing uses of tools within Science Applied. Whereas most financial research employed Reinforcement Learning in an attempt (often naïve) to let machines autonomously learn how to trade, we proposed a conceptual inversion: using Reinforcement Learning to understand the internal “brain” of a trading algorithm and to guide its calibration (a form of regularization) based on the “brain” (expert’s heuristics) of a trader.\\

More than anything else we wanted this to constitute a wake-up call in several domains, most notably \textbf{labour economics}, as we anticipated widespread - and unjustified - layoffs triggered by a misunderstanding of the proper interplay between human and machine intelligence. Interestingly, this concern mirrors the dynamics unfolding today\footnote{In our literature, we distinguish between two forms of efficiency, denoted as Efficiency A and Efficiency B. Efficiency B focuses on reducing operational \textbf{burden} - modifying processes to achieve the same level of output, quality, or service with fewer resources. While valuable, we argue that Efficiency B does not constitute the backbone of competition driven by innovation. By contrast, Efficiency A is more \textbf{ambitious}, concerned with creating substantially more value with the same resource base. This latter form of efficiency is considerably more challenging to achieve, yet far more impactful, and it is the form we explicitly advocate throughout our work.}.\\

As the consequences of \textbf{conflating Applied Science with Science Applied} continued to manifest, we published \cite{Advances in AI} in 2023 to clarify the dangers of this confusion, especially in the context of transformation. We argue that a substantial share of AI project failures arises because \textbf{methodologies are inherited from Science, while KPIs are measured by business standards}.\\

\subsection{Conclusions}
\label{subsection: Conclusions 2}

Compared to the 2025 Nobel laureates (Mokyr, Aghion, and Howitt), who studied the consequences of innovation at the macroeconomic level, \textbf{Algorithmization} - emerging from a decade-long synthesis of academic theory and industrial results - provides the \textbf{mechanisms that trigger innovation and propagate it across business layers}. While their work explains how technological progress shapes long-run growth, productivity, and creative destruction, Algorithmization focuses on \textbf{how technology must be architected and operationalized for those theoretical effects to materialize in practice}. It addresses the missing link between innovation as an abstract input and innovation as a repeatable, efficient, organization-wide process.\\

By redefining the firm as a \textbf{federated algorithmic ecosystem}, Algorithmization offers a framework in which efficiency, adaptability, and applied scientific reasoning scale coherently across products, departments, companies, sectors, countries and societies.\\

In this sense, Algorithmization does not simply describe a technological evolution; it constitutes a \textbf{discipline of Applied Science} that operationalizes innovation itself. It transforms abstract models into production-grade, federated architectures capable of evolving continuously - thereby unlocking the conditions under which Mokyr’s, Aghion’s, and Howitt’s macroeconomic mechanisms can actually play out in the real world. As a result, \textbf{Algorithmization could become a foundational pillar for the next era of economic organization}: one in which efficiency, interoperability, and strategic differentiation emerge not from isolated products or models but from the cohesive design of the entire machine behind them.\\

\section{Theoretical Framework}
\label{section: Theoretical Framework}
Humans require clear and precise definitions as a precondition for meaningful discourse. Without such precision, terminology becomes conflated with the underlying concepts.\\

This chapter brings to the foreground academic, social and business dynamics that, we believe, have created a perfect storm - an environment highly conducive to conceptual inefficiencies, misinterpretations, and widespread confusion.\\

From academic language dynamics and respected talent subject to constructing wrong narratives upon previous misleading ones, to the financial markets acting as a counterweight that restores alignment, this sequence illustrates a mechanism through which the system can ultimately converge back toward equilibrium.\\

\subsection{The Unintended Marketing Deviation in Contemporary Academic Language}
\label{subsection: The Unintended Marketing Deviation in Contemporary Academic Language}

Precise terminology is strictly followed in academic settings up to the point where terminology leads to complex or unwieldy nomenclature. To facilitate communication, scholars sometimes go back to basics and replace these terms with more eloquent or rhetorically elegant labels - e.g. Avatar Calibration in \cite{PhD}.

\textbf{The difficulty arises when such simplified labels - intended strictly for academic convenience - escape into broader public use}.

When this diffusion occurs at scale - as it has in the case of AI - the flow of information can even reach researchers through social rather than academic channels. This \textbf{inversion researcher-consumer} destabilizes the typical knowledge hierarchy and introduces novel forms of bias and confusion. 

The effect is particularly pronounced among younger audiences, who constitute the fastest-growing demographic engaging with AI in a world where coding and computational skills have been strategically promoted at national and international levels. When they \textbf{grow as scholars while biased from inception by social confusion}, it becomes particularly difficult for them to escape from the common myopia.\\

\subsection{Terminology: A Contextual Framework for Conceptual Clarity}
\label{subsection: Terminology: A Contextual Framework for Conceptual Clarity}
Before proceeding, it is essential to establish definitions so that the arguments developed in the remainder of this paper can be interpreted correctly.\\

For the purposes of this paper, we define \textbf{Artificial Intelligence (AI)} as \textit{Machine Learning supplemented by marketing narratives}; to date, no existing algorithm embodies intelligence in any substantive sense.\\

We define \textbf{intelligence} as \textit{the capacity to generate out-of-sample outliers with precision}.\\

Correspondingly, we define \textbf{Machine Learning (ML)} as a branch of \textbf{Computational Statistics}, thereby situating it conceptually much closer to Traditional Statistics than commonly acknowledged.\\

In both Traditional and Computational Statistics, models generally take the form \textbf{\begin{math} y = f(x) + u\end{math}}, where (\textit{y, x}) denote data points. Typically, the researcher wants to estimate \textit{y} through the nature of its relationship with \textit{x}, summarized by \begin{math} f(·)\end{math}, which represents the model. Finally, \textit{u} denotes the error term (see \cite{Advances in AI} for a deeper discussion). The principal distinction between the two paradigms lies in how \begin{math} f(·)\end{math} is constructed: Traditional Statistics largely relies on closed-form expressions derived from explicit mathematical assumptions and theorems, whereas Computational Statistics relaxes these assumptions and instead obtains the model through iterative numerical convergence upon a series of trial-and-errors.\\

It is important to recall that the function \begin{math} f(·)\end{math} fundamentally revolves around the leverage of averages. When confronted with unknown or random variables, scholars have consistently sought more refined ways to characterize their distributions and to use their expected values as one of the most rational basis for prediction. Techniques such as dummy variables (and their computational analogue, neurons, as discussed in \cite{Advances in AI}) or approaches grounded in Bayesian theory simply provide different modeling choices depending on the structure and nature of the data.\\

Quantum computing, on this line and given its complexity, remains at its early stage. This is, attempting to reproduce even the most elementary forms of \begin{math} f(·)\end{math} - albeit at unprecedented computational speeds. The distinction is crucial: so far quantum architectures are not yielding \textit{smarter} models but \textit{faster} execution of existing ones. As this paper emphasizes, achieving genuinely deeper algorithms depends on advances in the \textbf{Machine software’s flexibility}, not merely on accelerating computation. For this reason, we anticipate that the flexibility required for corporate innovation will remain largely beyond the quantum domain for decades.\\

It is also worth noting that the primary \textbf{barrier-to-entry for Traditional Statistics is mathematical judgment, whereas for Computational Statistics it is largely computational budget} for the aforementioned iterative optimization. Both paradigms are powerful and complementary; each offers distinct advantages depending on the nature of the problem. However, we argue that their most striking divergence today is not real impact but linguistic expectation - namely, the terminology used to describe analogous ideas. To illustrate, terms that are nearly equivalent in meaning have diverged substantially in jargon across fields:\\

\begin{figure}[ht]
\vskip 0.1in
\begin{center}
\centerline{\includegraphics[width=\columnwidth]{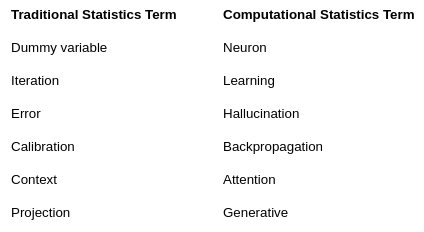}}
\caption{Illustrative cases of conceptual proximity, expectational divergence.}
\label{fig: Table Trad vs Comp}
\end{center}
\vskip -0.2in
\end{figure}

This linguistic drift has contributed substantially to public misunderstanding and to the overextension of claims about what contemporary systems can do. \\

Furthermore, it is important to recognize that the \textbf{scientific mindsets} underlying the two approaches diverge markedly. Broadly speaking, the former tends to be \textbf{theoretical}, proceeding step by step with formally proven results, whereas the latter is predominantly \textbf{experimental}, with outcomes still to be fully understood. The former emphasizes \textbf{ex-ante understanding}, while the latter relies heavily on \textbf{ex-post analysis - typically, trying to understand black-boxes by their behavior}. Traditional statistical models are typically \textbf{regularized}, imposing structural constraints to keep behavior within expected bounds, whereas many modern computational models are more overfitted, tethered to historical data in ways that make their behavior highly unstable when confronted with new inputs. Likewise, as suggested above, the former tradition is generally \textbf{transparent and auditable}, with models publishable in closed-form papers, while the latter often yields \textbf{black-box systems}, dispersed across extensive and frequently unstructured codebases.\\

These characteristics - reinforced by inflated and imprecise jargon - contribute to widespread fears regarding the supposed ‘unknown reach’ of AI. The combination of attributes in the vector \textbf{[experimental, ex-post, overfitted, black-box]} around LLMs creates a collective myopia that obscures a simple but crucial point: despite its mystique, the underlying formal structure of these systems remains a variant of \textbf{\begin{math} y = f(x) + u\end{math}}.\\

We define the \textbf{Machine in Machine Learning} or \textbf{Learning-Adaptive Machine} (see title in \cite{book}) as the layer of software responsible for operationalizing conditional logic - ‘if–then’ structures - within hardware environments.\\

This software layer gives rise to \textbf{Algorithms}, understood here as the synthesis of expert heuristics with tools from Computational and Traditional Statistics. Crucially, the performance of such algorithms is sensitive to the characteristics of the hardware on which they are deployed. For instance, in ultra-high-frequency trading, algorithms may exhibit markedly different performance across machines or physical locations due to the extreme latency sensitivity of this type of strategies.\\

When an algorithm assumes responsibility for autonomously generating its own \begin{math} ‘if–then’\end{math} logic, we refer to it as an \textbf{Autonomous Machine}. Such autonomy may arise through explicit instruction or through more subtle matricial mechanisms, such as the \textbf{Q-matrix} in Reinforcement Learning. While the former tends to be more flexible it is often slower than the latter.\\

Finally, we classify a Machine as \textbf{Federated} when its software is deployed across a distributed set of hardware endpoints - servers, personal computers, laptops, and similar devices - rather than residing as a monolithic block of code within a single server whose evolution typically requires hardware upgrades. This architectural shift carries several implications.\\

First, \textbf{Federated Machines are more cost-effective and environmentally sustainable}, as upgrades can be achieved more accurately and by recycling or repurposing existing hardware. Second, they \textbf{enhance intellectual-property protection by design}, a critical feature in organizations whose operational fabric increasingly resembles a large-scale algorithm. Third, they \textbf{enable organizational transformation to evolve in a federated manner}, combining the autonomy of decentralization with the synergy benefits of selective centralization. In contemporary competitive environments, these dual capabilities are essential: departments must adapt rapidly to peers, emerging cyber threats, regulatory changes, and talent rotation, all of which demand sustained reductions in the Time-to-Production (TTP) of strategic initiatives.\\

In practical terms, \textbf{Federation} comprises two fundamental layers:
\begin{enumerate}
    \item \textbf{Services}, which include shared components such as databases, messaging systems, and communication protocols.
    \item \textbf{Smart Agents (or Smart Nodes)}, which represent the modular units into which the system’s \begin{math} ‘if–then’\end{math} logic has been decomposed. Each node autonomously operates as an algorithmic entity capable of combining heuristics with statistical models - computational or traditional - analogous to the architecture of algorithmic trading strategies.
\end{enumerate}

In this framework, the term \textbf{Agentic AI}\textbf{ refers to the generation and orchestration of code through federated and synergetic Smart Agents}. It thus denotes a \textbf{distributed, node-based architecture} in which services and applications emerge from the coordinated behaviour of \textbf{modular computational units}, rather than from the aforementioned monolithic, centrally orchestrated codebases.\\

More specifically, as outlined in \cite{Data MAPs}, these modular units take three complementary forms:
\begin{enumerate}
    \item \textbf{Minimal Architecture Units (MAUs)}, which contain the core computational rationale of the unit, the requisite data-access capabilities for its data-driven operations, and the communication interfaces necessary for its participation within the federated network.
    \item \textbf{Minimal Architecture Extensions (MAEs)}, which constitute direct functional augmentations of MAUs, expanding their operational reach while preserving structural cohesion and interoperability.
    \item \textbf{Minimal Architecture Patterns (MAPs)}, which are higher-order orchestrations of MAUs and MAEs. MAPs generate full services and applications and are fundamentally grounded in \textbf{interoperability-by-design} across agents. Their performance is further enhanced through coordinated \textbf{software and hardware synergies}, likewise embedded into the architecture from inception - i.e. they are synergistic Custom SaaS.
\end{enumerate}

Within this framework, we distinguish between two types of agents:
\begin{itemize}
    \item When nodes incorporate explicit algorithmic structure, we refer to them as \textbf{Strategies-based AI Agents}.
    \item When a node consists largely of a statistical model and that model is an LLM, we classify it as an \textbf{LLM-based AI Agents}.
\end{itemize}

Accordingly, it is important to emphasize that Strategies-based AI Agents are not only compatible with LLM-based AI Agents; they can aggregate different LLMs. The opposite does not necessarily hold.\\

The remainder of this paper is intended to equip the reader with the conceptual and empirical basis needed to assess this claim: that, when the objective is scalable, resilient, and production-grade transformation, \textbf{Strategies-based architectures dominate LLM-based approaches}. Our analysis aims not to impose a conclusion, but to provide the framework through which an informed judgment may be reached.\\

\subsection{The Narrative Amplification is Largely not Academic}
\label{subsection: The narrative amplification is largely not academic}
Journalists, investors, civil servants, politicians - particularly in a context where geostrategic considerations have intensified, as discussed in \cite{EEN} - as well as entrepreneurs and corporate managers have all embraced these narratives. The \textbf{reputational strength of the brands promoting such claims} has functioned as a de facto quality certification, enabling their rapid dissemination despite the often ambiguous and weakly substantiated propositions underlying them.\\

Compounding the issue, many \textbf{social-media influencers lack the requisite technical background} and frequently amplify these narratives by leveraging the widespread human bias toward fear of missing out (\textbf{FOMO}). As noted earlier, the convergence of these factors has produced a socio-technical environment without historical precedent.\\

As a result, highly capable individuals across a wide range of influential positions have begun to echo - or even elaborate upon - incorrect statements. Many of these \textbf{actors previously served as trusted gatekeepers of expertise, yet they now inadvertently propagate misinformation}. This dynamic profoundly disrupts the social and industrial ecosystems that depend on accurate knowledge transmission.\\

\subsection{The Hardest-to-Persuade Talent as a Stabilizing Force}
\label{subsection: The hardest-to-persuade talent as a stabilizing force}

However, there exists a tipping point at which narratives crystallize into capital allocation. At that moment, accuracy and critical challenge become decisive sources of edge. \textbf{The discussion then enters the domain of the investment industry} - where one may not only take long positions on genuine innovation but also short positions when one disagrees with its inflation.\\

The current wave of hype in private markets is now spilling over into public markets. Yet these markets have long been exposed to the domain of \textbf{Algorithmic Trading} - arguably one of the most complex and inherently digital - native scientific challenges. \textbf{They are therefore acutely aware of how difficult it is to execute Applied Science correctly at scale}. Historically, only a select group of firms - such as Renaissance Technologies, Citadel, D. E. Shaw, and, more recently, XTX - have managed to succeed, and even then with an exceptionally narrow and disciplined focus. Expanding beyond this ‘laser-focused’ model is notoriously challenging (see \cite{book}, \cite{Inside the blackbox}, \cite{Advances in ML} for further discussion).\\

As a consequence, professional public-market investors are increasingly questioning the capacity of LLM-centric companies to generate sustainable profits. Because expected future activity now appears significantly lower than the projections promoted in private markets, the revenue outlook for the companies supplying the underlying AI chips is likewise being reassessed. It is important to emphasize, however, that what is being scrutinized is \textbf{not an AI bubble in general, but rather a more specific and contained LLM bubble}. We hope this paper also helps clarify the discussion and its market consequences.\\

The immediate outcome has been a \textbf{noticeable strategic business pivot}: many of these companies are \textbf{inorganically shifting from B2C markets} - where users readily adopt free services but show limited willingness to pay - \textbf{toward B2B markets}, where monetization prospects appear more viable.\\

In light of the foregoing considerations, we maintain that although the overall direction of this strategic pivot may be justified - particularly in view of the recent shift among CEOs toward impactful transformation rather than the superficial \textit{digitalwash} previously prevalent, as analyzed in \cite{Digital Washing} - the \textbf{judgment} guiding this shift and the \textbf{tools} selected to implement it remain fundamentally misaligned. We argue that meaningful, \textbf{scalable B2B transformation necessitates the adoption of Strategies-based Agentic AI}, as defined earlier and examined in greater depth in \cite{Data MAPs}.\\

In the sections that follow, we examine the challenges inherent in this strategic business pivot and elucidate its implications for the corporate strategy documents that are already shaping the trajectory of many major firms over the coming years.\\

\subsection{Conclusions}
\label{subsection: Conclusions 3}

The chapter establishes that \textbf{conceptual clarity is not a luxury but a prerequisite} for sound strategy. The linguistic ambiguity surrounding AI, the conflation of Applied Science with Science Applied, the misunderstanding of Machine versus Learning, and the erosion of credible expertise together form a perfect storm of misjudgment. Against this backdrop, the theoretical architecture developed here - grounded in Algorithmization, federated Machines, and Strategies-based Agentic AI - helps provide the analytical foundation needed to navigate the next stages of transformation with precision rather than hype.\\

The subsequent sections of this paper build upon this framework to examine the practical consequences for corporate strategy, technological design, and economic equilibrium.\\

\section{Why AI has Largerly Failed so Far}
\label{section: Why AI has Largerly Failed so Far}
We argue that the conclusion presented in MIT’s report\footnote{\href{https://www.artificialintelligence-news.com/wp-content/uploads/2025/08/ai_report_2025.pdf}{State of AI in business 2025.}} - namely, that approximately 95\% of AI projects fail - stems from a set of structural causes that can be grouped into four broad categories.\\

As we will demonstrate, these sources of failure can be traced back to a fundamental semantic confusion: as thoroughly discussed in \cite{Advances in AI}, \textbf{Science is not Science Applied, and neither is equivalent to Applied Science}.\\

\subsection{Companies are not Algorithmic Ecosystems}
\label{subsection: Companies are not Algorithmic Ecosystems}

Most companies are not designed as \textbf{algorithmic ecosystems} but a stack of technology and operational legacies. Their \textbf{algorithms and models are typically forced onto legacy operational structures rather} than integrated into a coherent, platform-native architecture. As a result, their behavior is neither smooth nor clockwork-like, and they fail to exhibit the characteristics of algorithmic-native or on-platform organizations.\\

\begin{figure}[ht]
\vskip 0.1in
\begin{center}
\centerline{\includegraphics[width=\columnwidth]{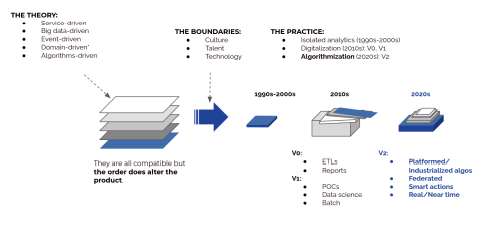}}
\caption{Technology stack evolution as predicted in \cite{Data MAPs}.}
\label{fig: tech stack evolution}
\end{center}
\vskip -0.2in
\end{figure}

Dimensions such as advanced permissions, responsibility allocations, regulatory constraints, ISO requirements, and internal politics constitute an ensemble of interdependent components that are far too numerous and heterogeneous to align naturally. Under such conditions, \textbf{systemic misalignment between Science Applied and Applied Science is not merely likely but structurally inevitable}. As a consequence, these arrangements are \textbf{bound to fail - and, in practice, they do fail}.\\

On this note, \cite{tesis ICADE} identifies organizational issues as the main driver in the AI failure while following a Technology-Organization-Environment (TOE) framework - well above the other two. This, we argue, evidences the need to work on letting organizations become on-platform by design as thoroughly explained in \cite{Data MAPs} and revisited in \cite{ASI}, \cite{3layer}, \cite{Advances in AI}, \cite{Modern Cybersecurity} and \cite{Vertical Integration}, \cite{EEN} among others.\\

In summary, \textbf{before an organization can meaningfully onboard so-called AI products, it must first undergo its own transformation process}.\\

\subsection{Providers are not Ready nor is their Clients’ Procurement}
\label{subsection: Providers are not Ready nor is their Clients’ Procurement}

On the supply side, many providers have simply \textbf{repackaged existing offerings} to appear consistent with the dominant AI narrative. Because genuine expertise in this field requires years of multidisciplinary training, leaders often remain insufficiently qualified to guide such efforts. As a result, their value propositions tend to be technically inaccurate, and their perceived leadership credibility erodes. This erosion, in turn, undermines their ability to attract and retain high-quality talent. Moreover, this dynamic increasingly affects the quality perception from the most advanced AI experts on the demand side - whose numbers, by design and necessity, grow over time - further widening the expertise gap between what organizations require and what providers are capable of offering.\\

On the demand side, Procurement departments routinely issue requests for quotes (RFQs) that subtly yet \textbf{systematically exclude solutions containing proprietary IP}. By requiring providers to fully disclose and document how they would solve the challenge, these RFQs generate an immediate \textbf{adverse selection} effect: overall, any serious innovator with meaningful IP is disincentivized from participating. Compounding this issue, many IT departments request \textbf{open-source access} to any proposed solution, a requirement even more detrimental to innovation than traditional RFQ practices. Moreover, when external solutions are coerced into internal, bespoke formats, their maintenance and evolution become disproportionately costly and operationally risky (see Figure \ref{fig: paper CEO}). Together, these mechanisms create a \textbf{structural paradox that prevents impactful innovation from entering the organization}.\\

Further, the post-project \textbf{KPIs} promoted by the first group and (often naïvely) embraced by the second are, in most cases, \textbf{structurally flawed projections}. In transformation - as in algorithmics more broadly - the final outcome depends on a high-dimensional set of variables. Two firms operating in the same sector and of comparable size may extract entirely different levels of value from the very same technological onboarding. There is a pervasive KPI-myopia that materially distorts both decision-making and provider selection. These numerical projections create a false sense of precision while ignoring the contextual, architectural, organizational, and talent-related factors that actually determine success. \textbf{Projecting them with precision is a deeply complex exercise} - one that requires rigorous architectural, organizational, and microeconomic analysis, not a pre-sales finger-in-the-air. Transformation outcomes cannot be reduced to simplistic extrapolations: they must be derived from a holistic assessment of the company’s workflows, legacy constraints, regulatory environment, talent structure, and strategic posture. We believe any KPI projection communicated without this depth of analysis is, by design, misleading. Moreover, such KPIs are typically derived from proprietary information obtained in prior client engagements. \textbf{Their public reuse not only violates the contextual specificity of those results but also undermines the competitive-advantage logic} that should govern transformation outcomes.

\subsection{Talent is not Properly Managed Cross the Roles in Transformation}
\label{subsection: Talent is not Properly Managed cross the Roles in Transformation}

Modern transformation is fundamentally a game of talent. As said, the global competition for individuals capable of driving algorithmic, AI-native change is so intense that leading technology companies - including Meta - are bidding hundreds of millions of dollars, and in some cases \textbf{more than a billion, for top-tier talent}. This is the scale of value attributed to the people who can truly unlock transformation.\\

In this context, \textbf{misassigning roles inside the company} - especially confusing Auditors with Catalyzers, as we shall see below - is not a minor organizational mistake. It \textbf{directly undermines the company’s future competitiveness}.\\
\begin{itemize}
    \item A true \textbf{Catalyzer} brings the vision, the methods, the knowledge and the architectural understanding needed to move the company forward.
    \item A right \textbf{Champion} is an expert intrapreneur, willing to take risks to chase returns.
    \item An \textbf{Auditor}, by design, observes and reports; they do not create or drive transformation. But they are smart assets that the senior management leverages.
\end{itemize}

Confusing them leads to stalled initiatives, diluted impact, internal (and external) friction, and, ultimately, strategic failure.\\

When senior management assigns trusted employees to transformation roles primarily to audit internal efforts - yet \textbf{fails to decouple them from catalyzer responsibilities} for which they lack the requisite interdisciplinary expertise - the organization exposes itself to \textbf{one of the costliest strategic errors available}. This misalignment of role, skill, and mandate systematically slows transformation, distorts technology-adoption pathways, and ultimately endangers the firm’s long-term competitive position.Notably, by late 2025 many firms began actively seeking strategic partnerships with organizations possessing the requisite attributes to \textbf{externalize this role} - an evolution again fully consistent with the dynamics anticipated in \cite{Digital Washing}.\\

\subsection{Science Applied has been Confused with Applied Science}
\label{subsection: Science Applied has been Confused with Applied Science}

As noted earlier, \textbf{Science} is fundamentally concerned with defining the function \begin{math} f(·)\end{math} - that is, with designing, validating, and refining models. Scholars dedicate significant effort to creating new methodologies or improving existing ones, and the scientific method, supported by its institutional community, generally performs well in advancing this frontier.\\

However, once those models leave the scientific domain and are applied to diverse real-world datasets, \textbf{Science Applied}, they are typically handled without sufficient domain knowledge to address the underlying business challenge. This mismatch - between the skills of the modeler and the needs of the domain - creates structural failure modes. A detailed discussion is presented in \cite{Advances in AI}, which can be summarized as follows:\\

\begin{figure}[ht]
\vskip 0.1in
\begin{center}
\centerline{\includegraphics[width=\columnwidth]{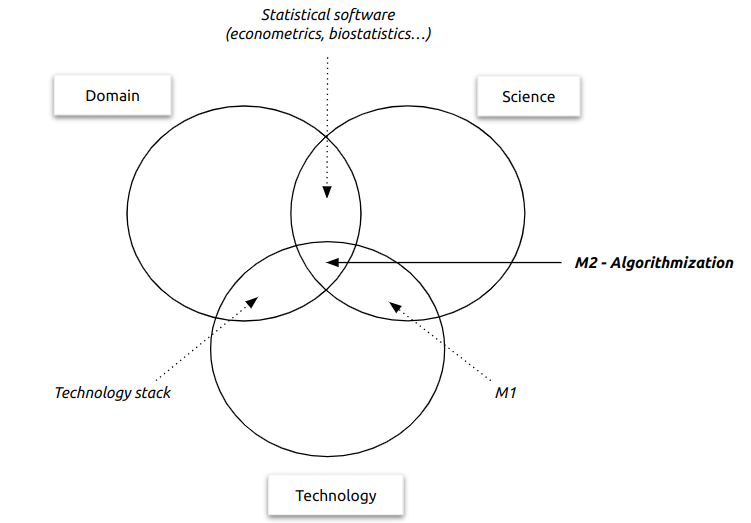}}
\caption{Schematic representation of the fields that are currently driving the Algorithmization process.}
\label{fig: 3circles M1M2}
\end{center}
\vskip -0.2in
\end{figure}

In real-world environments, a single model does not constitute a complete solution. Effective solutions, \textbf{Applied Science}, require the integration of \textbf{expert heuristics, computational statistics, and traditional statistical methods} - together forming what we have defined as an \textbf{algorithm}. Moreover, in competitive settings, solutions are inherently dynamic: they require the continual incorporation of new heuristics, updated models, and refined operational procedures. This iterative evolution is effectively endless. We therefore contend that creativity - namely, the capacity to generate novel heuristics and strategic insights - will remain the cornerstone of future human work, even as algorithmic systems become more pervasive.\\

\textbf{Transformation must be understood as a continuous process}, requiring ongoing adaptation as new technologies are unlocked. This dynamic is often more awkward and less visible in white-collar environments than in operational or factory settings. In production contexts, the impact of innovation is typically immediate and tangible. For example, when Omega - the Swiss watch manufacturer - introduced the \textbf{Co-Axial escapement}, invented by George Daniels, to reduce friction in the movement (one of the company’s key performance metrics), the firm subsequently reconfigured its entire production process. The technological improvement was \textbf{evident, measurable (yet not in terms of business impact), and therefore rapidly integrated} - illustrating how smoothly transformation can occur when operational constraints make the benefits unmistakable. How to design and implement white-collar factories was deeply analyzed in \cite{Data MAPs} - that is, organizational environments capable of bringing the discipline, repeatability, and operational clarity of industrial production into knowledge-based domains.\\

In summary, for the last decade technology of Science Applied has been awkwardly forced into business transformation which is a far more complex challenge.\\

\subsection{Conclusions}
\label{subsection: Conclusions 4}
\textbf{The 95\% failure rate is not a reflection of AI’s inadequacy, but of institutional misalignment driven by conceptual misunderstanding}.\\

This conclusion sets the stage for the remainder of the paper, which examines the architectural solutions - particularly M2 and Strategies-based Agentic AI - required to reverse these dynamics and enable transformation that is scalable, resilient, and economically meaningful.\\

\section{The Machine Theory of Agentic AI: M1 and M2}
\label{section: The Machine Theory of Agentic AI: M1 and M2}

After a decade at the forefront of advanced transformation initiatives, \textbf{we have identified a foundational structure} that should be taken into account when discussing innovation in algorithmics and technology.\\

From this point forward, we encourage the reader to \textbf{distinguish explicitly between the Machine and the Learning components of Machine Learning}, as they represent two fundamentally different challenges with distinct scientific and operational implications.\\

The Machine can be further decomposed into two distinct components:
\begin{enumerate}
    \item \textbf{M1}, which comprises the mechanisms required to \textbf{\textit{build} the Learning component (the L)} - that is, to construct, estimate, and expose the statistical model \begin{math} f(·)\end{math}; and
    \item \textbf{M2}, which comprises the mechanisms required to \textbf{\textit{consume} the L}, integrating it into algorithmic, strategic, and operational architectures across the enterprise.
\end{enumerate}

As we will show, \textbf{consuming Ls is substantially more complex than producing them}. While M1 focuses on generating \begin{math} f(·)\end{math}, M2 must embed this function within federated, resilient, and strategically coherent systems - an endeavor that introduces architectural, organizational, and algorithmic challenges far exceeding those of model construction.\\

\subsection{The L in ML}
\label{subsection: The L in ML}

The \textbf{Learning} component of Machine Learning - the \textit{L} in ML - corresponds to the combined domain of \textbf{Computational Statistics} and \textbf{Traditional Statistics}.

From a transformation perspective, this component represents the \textbf{Science}: the model - \begin{math} f(·)\end{math} - is its central challenge. The field has evolved over decades through sustained academic research aimed at advancing the frontier of model development. As a result, \textbf{academia retains a structural advantage} in this domain.

Because the academic community is large, globally distributed, and grounded in open scientific exchange, corporations generally cannot compete in this domain. Indeed, one may argue that in industrial settings \textbf{\begin{math} f(·)\end{math} ultimately becomes a commodity}, as no single firm can sustain a competitive advantage while a vast research ecosystem continues to push the model frontier forward. 

Thus, we decided to shift our research focus from the \textit{Learning} component to the \textit{Machine} component of Machine Learning, a transition undertaken through our Centre of Excellence since 2015. This is, \textbf{from the capacity to evolve the theory to the capacity to translate theory into practice}.\\

\subsection{The M1 in ML}
\label{subsection: The M1 in ML}
The first \textbf{Machine} in Machine Learning - \textbf{M1} - is centered on the \textbf{calibration of the scientific \begin{math} f(·)\end{math}} through the integration of sophisticated software engineering and hardware management.\\

This is, M1 encompasses the software and hardware mechanisms required to:

\begin{enumerate}
    \item \textbf{Transform data into numerical representations} suitable for statistical estimation (e.g., tokenization, embeddings, tensors).
    \item \textbf{Execute large-scale iterative optimization} to estimate the parameters of \begin{math} f(·)\end{math} (e.g., gradient descent, backpropagation, distributed training).
    \item \textbf{Manage the computational environment} - GPU clusters, distributed systems, memory management - that enables such optimization to occur at scale.
    \item \textbf{Deploy and serve the resulting model} to downstream applications or user interfaces.
\end{enumerate}

In other words, \textbf{M1 operationalizes the Learning component (the L) in Machine Learning} by providing the machinery required to estimate and expose \begin{math} f(·)\end{math}. Its primary output is a model whose design and improvement are largely driven by global academia.\\

As said, we refer to this challenge as \textbf{Science Applied} (see \cite{Advances in AI}) - this is, a linking stage between Science and Applied Science. It has become a major source of industry confusion because many of the practices that are optimal for advancing science are poorly suited to producing realistic, high-impact solutions. As noted throughout this document, these limitations include the absence of domain heuristics, the reliance on incremental improvements to a single model, and the assumption that model evolution equates to solution evolution.\\

\begin{figure}[ht]
\vskip 0.1in
\begin{center}
\centerline{\includegraphics[width=\columnwidth]{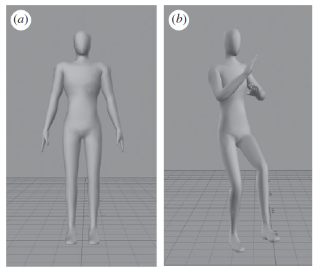}}
\caption{Snapshots from paper \cite{Avatar dance} which we often use as an eloquent example of Science Applied.}
\label{fig: Male dance}
\end{center}
\vskip -0.2in
\end{figure}

We further contend that \textbf{the confusion between Science and Science Applied extends into academia itself}. Several prominent awards have recently been granted to authors who applied existing models - this is, where \begin{math} f(·)\end{math} was already a commodity - using data and conceptual insights developed by domain experts, and who primarily contributed advanced data-engineering pipelines to calibrate these models. To be clear, M1 is a legitimate and challenging domain, and excellence in M1 can indeed merit recognition. However, such recognition should focus on the M1 contribution itself, not be conflated with the creation or advancement of \begin{math} f(·)\end{math}.\\

\subsubsection{LLM-based Companies as M1}
\label{subsubsection: LLM-based Companies as M1}

The most popular form of M1 today is the infrastructure supporting \textbf{Large Language Models (LLMs)}:
\begin{itemize}
    \item \textbf{Foundational model}: The pure implementation of \begin{math} f(·)\end{math} that is shared - almost identically - across numerous companies. These models are extremely costly to train, deploy, and fine-tune. This is, \textbf{budget is their major barrier to entry - well above their academic sophistication}.\\
    Eloquently, an LLM converts text (prompts) into numeric tokens, applies \begin{math} f(·)\end{math} to generate a projection - once more, a numeric estimate of y in the expression \begin{math} y = f(x) + u\end{math} - and then converts these projections back into text resulting in a \textbf{chatbot-like experience by design}. This is, there is nothing really new in the way researchers create models beyond the (initial global surprise and) user experience of the person interacting with the model. Since the database largely corresponds to information extracted from the internet, users can progressively explore a topic by exploiting linear or even convex combinations of existing knowledge. The result is an output with the \textbf{average accuracy} (often confused with intelligence) of those who wrote about the topic - notably, average within \textbf{information often curated by experts (from communities, journals…)}.
    \item \textbf{A new bar yet to understand}: Thus, because LLMs efficiently aggregate these sources, that average may appear to perform at, say, ‘7–8’ relative to individual non-experts. As shown in \cite{MIT brain}, the cognitive load of an LLM user is significantly reduced: the user must craft an effective prompt and evaluate an already-generated solution rather than produce the solution independently. However, this introduces an important shift: if LLMs deliver output equivalent to a ‘7–8,’ then this performance becomes the new human baseline to beat. \textbf{Professionals must therefore aim for the equivalent of a ‘12–13’ to maintain differentiation. In other words, the bar must rise}, and this rising bar will ultimately re-equilibrate cognitive activity in the human brain - figuring out how to go beyond the tool.
    \item \textbf{Beyond the LLM up to an app}: Extra layers surrounding the foundational model up to an app: so that users can interact with the model all over the world the companies based on LLM models need to overcome the challenges of any SaaS company. Some are standard from SaaS and others, more particular from the nature of LLMs. For example, from large-scale \textbf{parallelization} for millions of users and B2C-oriented UX features (voice interfaces, suggestion prompts) to lightweight graphical front ends.\\
    
    This layer begins to resemble initial components of \textbf{M2}, yet remains insufficient for B2B use. These systems still struggle with accuracy, high deployment costs (High-Performance Computers, HPC, required) due to inefficient memory usage (e.g., brute-force Docker deployments on increasingly expensive RAM), limited compliance due to black-box behavior, and incomplete understanding of how long-term and short-term memory affect model output. That may be the reason why agents are typically bound to data queries instead of code federation.
    \item \textbf{LLM data, not available on the internet}: It is important to recognize the \textbf{changing nature of data}: from publicly available internet data present in the calibration of the foundational model to privately generated in-session data. Historically, individuals contributing information to the internet exercised a degree of self-censorship, often auditing their own work to avoid exposing gaps in their knowledge; as a result, public data tends to exhibit a certain conservative quality. This is, it softly resembles properties of the peer-review from academic research - which is not systematically correct per paper but much more so over time and across papers. In contrast, \textbf{in-session data} - produced in real time through private interactions - faces no such social constraints and can therefore drift freely, sometimes generating signals that are orthogonal to what would have been expressed publicly. The \textbf{structural quality} of these two data sources is therefore fundamentally different. Any algorithm intended to operate on both must take this distinction seriously (exploitation of quality vs exploration of new discoveries) - yet, at present, \textbf{no meaningful advances have been made to explicitly account for this differentiation}.\\
    
    The cornerstone of this inefficiency - this over-reliance on LLM-based Agentic AI - lies in a pervasive confusion: namely, \textbf{the latest belief that AI is equivalent to LLMs, and that interacting with AI necessarily involves prompt-driven dialogue}. As established throughout this document, this view is incorrect and largely a product of commercial momentum rather than scientific reality - hence, it is prone to disappear in the mid run.
\end{itemize}

\subsubsection{The Most Complex Form of M1}
\label{subsubsection: The Most Complex Form of M1}

Other M1 implementations exist beyond LLMs - \textbf{any of those that largely rely on technology-intensive science to provide a service}.\\

Probably, the most popular instance before the irruption of LLMs, often seen as the pioneering one and still to date the most complex of all M1 for a number of reasons (including economic incentives), is \textbf{algorithmic trading}. Ultra-high-frequency trading systems require extremely complex, low-latency infrastructure, often located within the same physical building as the exchange to achieve the fastest possible message propagation - otherwise, the same algorithm could exhibit markedly different performance. These are \textbf{quintessential M1 systems}: tightly engineered, hardware-intensive, and focused on precise calibration.\\

Further, non-latency-sensitive algorithmic trading typically leverages time series market data. These appear homogeneous on the surface, yet they are generated by a \textbf{continually shifting set of distributions} shaped by market participation, sentiment, urgency, news flows, and geopolitics. \textbf{Hence, unlike the case of LLMs - where the \begin{math} f(·)\end{math} has effectively become a shared commodity - academia continues to search for the \begin{math} f(·)\end{math} capable of explaining and, ultimately, solving financial markets}. 

Still, flexible and interoperative strategic reasoning is not their target - which is the reason why an M2 is further required.\\

\subsection{The M2 in ML}
\label{subsection: The M2 in ML}

The second \textbf{Machine} in Machine Learning - \textbf{M2} - is \textbf{centered on the Algorithmization of the whole company upon new workflows for humans and bots} (up to their own self-calibration).\\

As said, we refer to this challenge as \textbf{Applied Science} (see \cite{Advances in AI}) and, \textbf{due to the complexity of its nature, we expect it to be the most proprietary of the two}.\\

As stated above, for the last two decades, as academics and professionals, we have observed that the  true target of advanced transformation is the corporate Algorithmization: the systematic construction of the \begin{math} 'if-then'\end{math} logic that enables organizations to integrate expert heuristics with \begin{math} f(·)\end{math} - including externally sourced Learning components - seamlessly across all operational areas. This is the path toward becoming a fully on-platform organization.\\

Examples across operational areas include:\\

\textbf{Lean IT}\\
We believe companies need to embrace a new type of architecture. In legacy \textbf{tech stacks}, new models do not naturally propagate across layers. Heterogeneous requirements - from ISO standards to cybersecurity protocols - create \textbf{friction points} that often prevent their effective onboarding. By contrast, a \textbf{tech} architecture that is federated, meshed, networked reduces these barriers by enabling smoother horizontal and vertical integration of algorithmic components.\\

On that note, Algorithmic infrastructures must be \textbf{continuously revisited} to reduce unnecessary complexity. \textbf{The leaner the architecture, the easier it becomes to introduce new code and unlock new operational flexibilities}. This continual simplification is essential for sustaining long-term adaptability. It also reinforces why we have concentrated our efforts on this specific class of architectural challenges, while leaving consultants and internal data-science teams to focus on laser-focused, model-centric projects built around Ls, as discussed later in this paper.\\

A defining characteristic of Algorithmization is thus its \textbf{edge-computing nativeness} towards intellectual property protection, delivery flexibility and cost-consciousness (including environmental): it is not RAM-predatory, can execute efficiently across heterogeneous operating systems, and avoids the legacy constraints that burden centralized architectures.\\

Note that because M2 includes the ability to orchestrate hardware configurations when needed, \textbf{M1 is actually a subset of it}. And, thus M2’s conceptual scope is also significantly broader than that of M1. This is why \textbf{we often avoid adopting the traditional M1 jargon: it is narrow to capture the full scope of possibilities enabled by M2} and, for that reason, could become outdated within the next decade - much as we have long anticipated for the proliferation of data lakes, data swamps, and data meshes.\\

\textbf{Proprietary cybersecurity}\\

Algorithmization is also \textbf{critical for cybersecurity}. As discussed in \cite{Modern Cybersecurity}, cyber-attackers operate with minimal legacy constraints and are highly specialized in deploying innovative algorithms to intrude, exfiltrate, or disable systems for ransom. M2-level algorithmic infrastructures are therefore \textbf{indispensable defensive assets to unlock a timely response upon flexible, fast deployments that leverage Game Theory at their core}.\\

\textbf{Compliance by design}\\
Moreover, Algorithmization enables the efficient incorporation of transparent, auditable and traceable \textbf{compliance by design}, rather than as fragile after-the-fact add-ons.

\textbf{LLM-based latest attempts}\\
We are increasingly observing reports from firms that have deployed LLM-based agents as the core of their agentic infrastructures - particularly among large B2B organizations with significant IT budgets but relatively low regulatory constraints (e.g., in construction and adjacent sectors). These experiences indicate recurrent limitations: stochastic behavior intrinsic to probabilistic generation, high operating costs driven by reliance on high-performance computing, limited interoperability across agents, and challenges in meeting compliance and auditability requirements.\\

Taken together, these outcomes are consistent with the structural constraints of M1 - only approaches. They suggest that architectures centered exclusively on model-centric capabilities, without a corresponding M2 layer, are insufficient for scalable, resilient, and production-grade transformation.\\

\subsection{Conclusions}
\label{subsection: Conclusions 5}

\textbf{The Machine Theory of Agentic AI reframes the frontier of AI transformation}. The competitive terrain of the next decade will not be defined by who produces the most advanced \begin{math} f(·)\end{math}, but by who builds the architectures capable of \textit{consuming} these models effectively - architectures that are federated, strategic, interoperable, and inherently algorithmic.\\

\textbf{M1 creates the model; M2 creates the organization that can consume it}. It is in M2 - not M1 - where the structural, durable, and proprietary advantages of advanced transformation reside.\\

\begin{figure}[ht]
\vskip 0.1in
\begin{center}
\centerline{\includegraphics[width=\columnwidth]{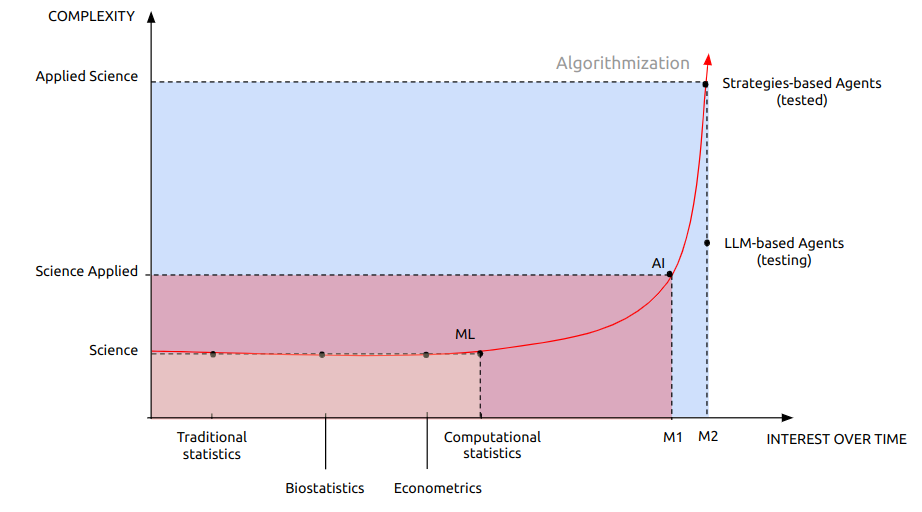}}
\caption{An schematic overview of the path across Science and Science Applied that has evolved into the current Applied Science discussion.}
\label{fig: Timeline}
\end{center}
\vskip -0.2in
\end{figure}

\section{Insights of the First M2}
\label{section: Insights of the First M2}
After a decade of developing and refining \textbf{M2}, this white paper represents our effort to synthesize and democratize the entirety of that experimental journey already discussed along our extensive literature on the novel field.\\

Furthermore, as noted above, we anticipate that this domain will become increasingly \textbf{secretive} across companies - much like Algorithmic Trading platforms, which are often characterized as \textbf{an arms race} (see \cite{Philip Treleaven}). For this reason, it is pertinent to shed light on the underlying mechanisms at this stage.\\

To facilitate a comprehensive understanding, we will incorporate our \textbf{professional experience} throughout the discussion, thereby providing the reader with detailed insights across all dimensions of the challenge.\\

\subsection{A Greenfield}
\label{subsection: A Greenfield}

\textbf{Long-horizon, disruptive innovation} is difficult to sustain within traditional corporate or academic environments. For this reason, we created our own \textbf{Centre of Excellence}, with a deliberately skewed composition - \textbf{99\% technology, 1\% business} - so that we could expand innovation while proving it in the industry at the same time that we bootstrapped the whole project.\\

Our aim was to establish a new discipline at the intersection of \textbf{Economics, Technology, and Machine Learning}, which we termed \textbf{Algorithmization}. Its purpose was to understand the mechanisms underlying \textbf{supply shifts induced by technological improvements}.\\ 

\begin{figure}[ht]
\vskip 0.1in
\begin{center}
\centerline{\includegraphics[width=\columnwidth]{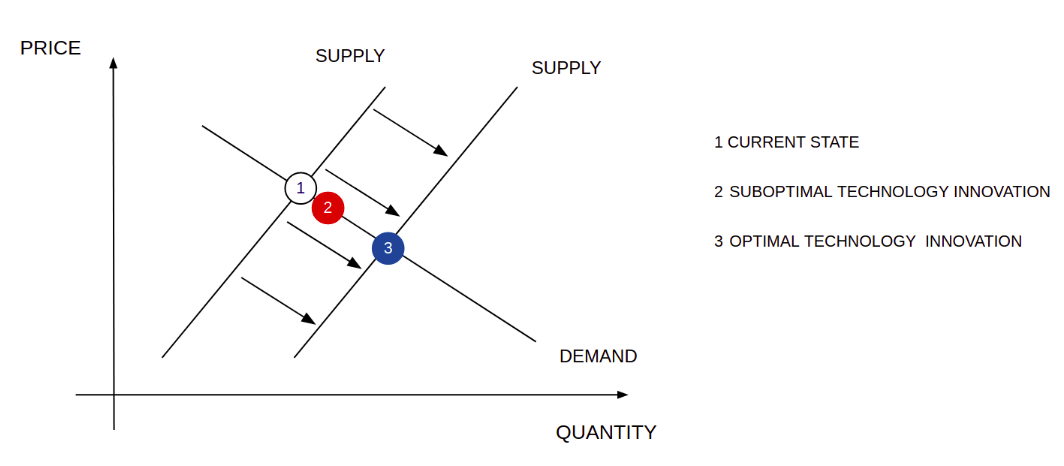}}
\caption{Schematic representation of the target state of Algorithmization is presented in \cite{Data MAPs}.}
\label{fig: Supply curve shift}
\end{center}
\vskip -0.2in
\end{figure}

In this sense, despite the surrounding Machine-Learning and technology jargon, this is fundamentally an \textbf{Economics paper}, one that required us to master (and in some cases pioneer) contributions in the other two domains in order to reach a genuine solution. Our background in Econometrics and Biostatistics proved essential in overcoming the barriers-to-entry.\\

As said, this convergence of disciplines - combined with the fact that our peers include both academics and industry practitioners - explains why much of our research is published on \textbf{SSRN}, as no Tier-1 journal yet specializes in this emerging domain.\\

Last, as previously noted, we anticipated that many readers would eventually rely on language models to fully comprehend our papers. Given their breadth and disruptive nature, we understood that acceptance by academia and industry would be gradual and, in many cases, ex-post. It is worth recalling that this expectation dates back to 2015 - long before the explosion of LLMs - when we foresaw that, as research topics grew ever more complex, the use of computational assistants might become standard practice in scholarly work.\\

With this in mind, and recognizing that our papers would increasingly be assessed or interpreted by machines, we reasoned that improving model accuracy (or, in contemporary terms, reducing hallucinations) required deliberate shaping of the underlying statistical distribution. Specifically, we sought to ensure that the distribution became sharply concentrated around each targeted sub-message, thereby limiting a model’s propensity to interpolate from unrelated regions of its learned space. We applied this principle not only in the design and structure of our papers - as discussed in \ref{subsubsection: Exploitation II: Market Fit} - but also in our internal project-management methodology, where each task carried a machine-oriented descriptive vector to facilitate precise computational interpretation.\\

\subsection{A Decade of Relentless Effort}
\label{subsection: A Decade of Relentless Effort}

\subsubsection{Seed: Algorithmic Trading as our M1}
\label{subsubsection: Seed: Algorithmic Trading as our M1}

As illustrated in \cite{Data MAPs}, our initial focus was \textbf{Algorithmic Trading}, which we consider one of the most complex Applied Science domains. This makes algorithmic trading a stress test of any machine-learning or algorithmic innovation.\\

\textbf{It took three years} to build a machine capable of matching - and then surpassing - the experimental sophistication described in \cite{PhD}. We realized \textbf{we had reached a pinnacle of autonomous learning} in non-latency-sensitive algorithmic trading when we completed a full \textbf{virtual-reality simulation environment}, including a custom-built exchange, broker, and data provider. This environment allowed researchers and compliance officers to examine algorithmic behaviour under both realistic calibration scenarios and extreme risk conditions. By enabling the machine to autonomously select scenarios 24/7, we had effectively created one of the \textbf{first autonomous AI machines}. Although the costs exceeded the return-on-assets required for production deployment (and we believe it to still be the case), the system remained ready for future stages. This work - along with additional risk-management designs in which supervisory algorithms could take control of trading algorithms - was selected by Oliver Wyman in 2019 to illustrate advanced risk-management concepts to the energy sector.\\

\subsubsection{Exploration before Exploitation: a Relentless Contrarian View that Proved to be Crucial}
\label{subsubsection: Exploration before Exploitation: a Relentless Contrarian View that Proved to be Crucial}

It is worth noting that a particular conversation proved pivotal in triggering the next stage of SciTheWorld. This took place with Jack Dorsey, founder of Twitter, to whom we were introduced during his visit to Spain to discuss Machine Learning.\\

Within Reinforcement Learning, we place considerable emphasis on optimizing the \textbf{exploration–exploitation} trade-off. We like the weighing of both worlds. So we not only used the concept as an algorithmic principle but also as a broader \textbf{innovation vs business philosophy} - being our equilibrium largely biased towards exploration.\\

After analysing the business models and product portfolios of major BigTech firms, we concluded that they tend to operate in the reverse regime: \textbf{exploitation over exploration}, driven more by commercial imperatives than by genuine innovation. When asked directly, Dorsey significantly corroborated this view.\\

We maintain that technology is far too complex to tolerate suboptimal design. Under such conditions, upgrades typically require either a complete rebuild or, more often, the accumulation of patches upon patches - resulting in the characteristic \textbf{spaghetti code} and architectural fragility. At the same time, we recognize the pressures faced by Silicon Valley entrepreneurs: venture capital expectations force an early demonstration of market fit, measured through \textbf{revenue-centric KPIs} rather than long-run P\&L.\\

Our \textbf{bootstrapped structure} therefore became a competitive advantage - \textbf{we could select exploration over exploitation}. Freed from external constraints, we were able to explore the full design space until the technology not only exceeded the client’s expectations in the short run (market fit) but was also aligned with her \textbf{future strategic needs}. In this way, we achieved a rare alignment of \textbf{short-run and long-run} objectives - even though the latter are inherently difficult to pitch.\\

\subsubsection{Abstraction: an M2 as a Generalization of our M1}
\label{subsubsection: Abstraction: an M2 as a Generalization of our M1}

The creation of a hedge fund was a major target of our CoE.\\

Once the trading and risk management technology were ready we subsequently considered launching it. This is, letting our M1 become a company.\\

We just needed to evaluate the technological infrastructure necessary to run the organization - namely, enterprise resource planners (ERPs). Our investigation revealed that existing ERPs were neither sufficiently \textbf{broad} nor sufficiently \textbf{algorithmic} as we had expected. So, we decided to do it ourselves with the same level of sophistication as the trading platform itself.\\

As discussed in \cite{Vertical Integration}, if one can design a complex strategy composed of multiple substrategies to respond to constantly shifting market conditions, then one can equally design strategies for \textbf{cybersecurity, HR, project management, server maintenance, procurement, and marketing}. The logic naturally generalizes, as the latter domains become \textbf{sub-challenges nested within the broader architecture} of the former.\\

It then took an \textbf{additional four years} (burdened by the COVID-19 period) to abstract the approach beyond trading, generalize it to all corporate departments, and project the resulting architecture back onto trading to validate correctness - an effort described in detail in \cite{Data MAPs} that gave rise to, we argue, the \textbf{first Agentic AI infrastructure for AI-enterprises and operations. The first design for the M2}.\\

It is important to note that, at this stage, our papers and demonstrations had begun to attract the attention of numerous advanced institutions. \textbf{We had a brand} (not to confuse with massive marketing) that competed with the main players. We were invited to advise the Bank of England on blockchain-related and digitalization initiatives; J.P. Morgan sought brief guidance on the transition from Science Applied to Applied Science; and Mubadala expressed interest in modernizing every dimension of its investment process - from private-equity operations to portfolio-company transformation and internal governance. Several additional engagements, protected under nondisclosure agreements, further reinforced the breadth of interest in our approach.\\

These engagements could easily have drawn our Centre of Excellence toward a more business-oriented posture - in other words, toward \textbf{exploitation} rather than exploration. However, we chose to remain fully committed to the \textbf{exploration phase} until we were confident that we had reached the \textbf{top of the sigmoid curve} in transformation - where large technology efforts would lead to marginally impactful innovation. By this we mean that we had developed all the essential tools required to address, in an integrated manner, the majority of future challenges we could plausibly anticipate.\\

\begin{figure}[ht]
\vskip 0.1in
\begin{center}
\centerline{\includegraphics[width=\columnwidth]{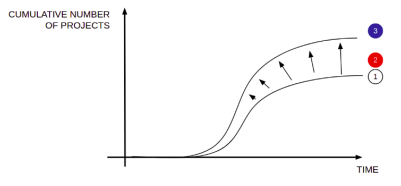}}
\caption{Sigmoid described in Data MAPs (see \cite{Data MAPs}).}
\label{fig: Sigmoid}
\end{center}
\vskip -0.2in
\end{figure}

\subsubsection{Hands On: Deeptech Developed at Unseen Speed}
\label{subsubsection: Hands On: Deeptech Developed at Unseen Speed}
\textbf{Within eighteen months}, we populated the M2. We had developed technology for most corporate departments that we could unlock, as a whole, to solve each departmental challenge - effectively an \textbf{AI-first ERP per department as we further developed our own methodology in parallel}.\\

Further, we deployed it compatible with any legacy ERP by allowing the production architecture (PA) of a company to be interoperatable by design via an extended production architecture (EPA).\\

\begin{figure}[ht]
\vskip 0.1in
\begin{center}
\centerline{\includegraphics[width=\columnwidth]{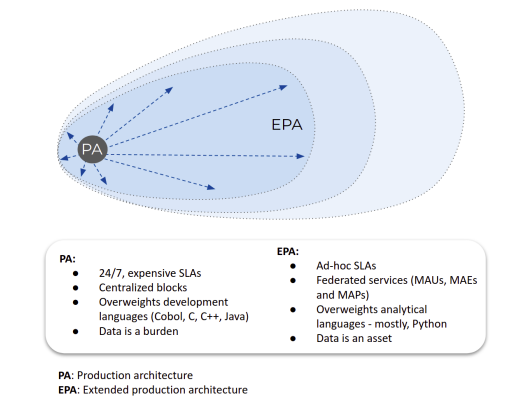}}
\caption{Extended-Production Architecture described in Data MAPs (see \cite{Data MAPs}).}
\label{fig: EPA}
\end{center}
\vskip -0.2in
\end{figure}

The efficiency with which we were able to evolve software was sufficiently notable that the Bank of Spain conducted a brief audit of our framework to assess its feasibility.\\

\textbf{We contend that this accelerated progress was driven not only by the agentic nature of our technology but, more subtly, by our decision to adopt Time-to-Production (TTP)} as the primary KPI. Traditional ROA metrics are poorly suited for evaluating experimental architectures, whereas TTP provided the most accurate measure of genuine transformational progress.\\

\subsubsection{Exploitation I: Internal Usage}
\label{subsubsection: Exploitation I: Internal Usage}

\textbf{The following eighteen months} were dedicated to iterative refinement. During that period, we enhanced server control mechanisms to extend \textbf{M1 capabilities beyond large tensor - and matrix-dependent computations, deploying additional locally resilient agents to increase robustness, lightness and flexibility} (at the price of speed on large databases).\\

As discussed in \cite{Modern Cybersecurity}, we place particular emphasis on \textbf{business continuity}, developing agentic strategies for both the CTO and the CISO that enable the infrastructure to \textbf{maintain and defend itself autonomously} as much as possible. Proprietary capabilities are key to ensure Game Theory can be unlocked and massive attacks avoided by design.\\

To bring the exploration phase to a close - an era intentionally and overwhelmingly biased toward technological capability rather than business application - we agreed to define the \textbf{top of the sigmoid curve} as the moment at which we could deploy \textbf{micro–smart agents directly on employees’ laptops}. This milestone ensured true \textbf{edge-computing capability}, maximal flexibility in user interaction, and a new standard of autonomous business continuity.\\

\subsubsection{Exploitation II: Market Fit}
\label{subsubsection: Exploitation II: Market Fit}

It is also relevant to note that, at that stage, the vision and mission statements of many BigTech companies had begun to converge toward LLM-based Agentic AI. In other words, the global state-of-the-art - at least as framed by the \textbf{technology industry - was effectively following} (and still testing) the path for what we had already accomplished. This dynamic naturally facilitated our marketing and commercial engagement, as prospective clients had already been primed by media narratives.\\

Despite this favourable environment, \textbf{we deliberately chose not to adopt LLM-based approaches} as the foundation of our agentic infrastructure. We remain unconvinced that \textbf{vibe coding} - the trial-and-error prompt-driven paradigm - is an optimal or sustainable path for industrial transformation. As argued in \cite{Digital Washing} - and subsequently confirmed three years later by OpenAI in \cite{Why_Hallucinate} - hallucinations are \textbf{structural features} of LLMs. When deployed in production environments, these structural errors introduce \textbf{unknown and inherently unpredictable failure modes}, which must be added to existing operational and cybersecurity risks. Under such conditions, relying on LLMs as core production systems does not appear to constitute a \textbf{responsible or robust architectural approach}. In a similar spirit, we minimized blockchain to its \textbf{organic, internal-use cases} and deferred quantum computing, which remains far too immature to function as an \textbf{M1}, let alone support the structural and strategic demands of \textbf{M2}.\\

Notably, as an informal robustness check, we examined how contemporary large language models interpret the conceptual structure presented in Chapter 3 of \cite{Data MAPs}. When provided with this material, several models (e.g., ChatGPT, Claude, Gemini) independently classified the approach as aligned with what is now described in the practitioner literature as “AI-first enterprises and operations”, consistent with definitions articulated in McKinsey's article\footnote{\href{https://www.mckinsey.com/featured-insights/mckinsey-explainers/what-is-an-ai-agent.}{What is an AI Agent?.}}. We make this experiment publicly reproducible to allow independent assessment. While such observations do not constitute validation, they suggest that the conceptual framing anticipates categories that later became formalized in industry discourse. The remaining chapters and the broader body of work extend this framework further, addressing dimensions not captured by these later classifications.\\

\subsubsection{A Priviledged Methodology Upon Two State-of-the-Art Platforms}
\label{subsubsection: A Priviledged Methodology Upon Two State-of-the-Art Platforms}
As a result of the developments outlined above, \textbf{we can now leverage two state-of-the-art platforms as foundations for subsequent waves of innovation}. Each addresses a different axis of transformation, and \textbf{together} they enable forms of organizational and financial innovation that were \textbf{previously unattainable}.\\

\textbf{Fractal}\\

Fractal unlocks the capacity to \textbf{onboard AI products while simultaneously triggering structural operations transformation across departments}. We conceptualize these transformation processes as \textit{bubbles}: initially independent, but ultimately converging into a single organizational architecture as they grow in depth and breadth.\\

To catalyze these bubbles, each department receives an all-in-one platform - \textbf{not as an end product, but as a starting point} - which is subsequently customized and evolved into a design that is interoperable across departments and compatible with the legacy by construction (see Data MAPs’ Extended Production Architectures in Figure \ref{fig: EPA}).\\

This breadth of departmental technology is essential: meaningful transformation is inherently more complex than simply onboarding AI products or performing Science Applied. Ultimately, \textbf{each head of department requires a technological infrastructure as sophisticated and supportive as that available to the CEO herself}. Fractal provides this operational leveling.\\

\textbf{AlphaDynamics}\\

AlphaDynamics unlocks modern capabilities in algorithmic trading and portfolio management. It democratizes the techniques traditionally reserved for expert traders and asset managers, enabling CFOs and heads of procurement to leverage the same level of analytical and strategic sophistication. This \textbf{closes a long-standing gap between financial markets expertise and corporate financial decision-making}.\\

\textbf{Combined impact}\\

The joint use of Fractal and AlphaDynamics enables entirely new modes of working - modes illustrated throughout the works mentioned in Literature Review. Together, they create a bidirectional bridge between corporate operations and financial strategy, allowing organizations to \textbf{function with unprecedented coherence, adaptability, and intelligence}.\\

\textbf{Scalability}\\

Scalability of \textbf{custom technology was a major breakthrough} for us.\\

Once achieved, our next objective was to scale the business itself. At the outset, persuading companies that they needed to undergo advanced transformation - particularly that indiscriminate data enablement was not always optimal for Algorithmization, and in some cases even counterproductive - was a difficult proposition. Today, however, these ideas are widely reflected in media discourse and academic publications. As a result, the \textbf{commercial motion can now be driven directly by us and/or amplified through partnerships} with consulting firms (whether strategy or technology focused) that wish to deliver their services - be it operational redesign or their own Ls - on top of our platform. Frequently, these “consultants” will in fact be the client’s own strategy units or data-science teams. In other words, it is a platform upon which any qualified actor can build transformation.\\

\subsubsection{Industry Recognition}
\label{subsubsection: Industry Recognition}
Industrially, the \textbf{Algorithmization methodology has been endorsed not only by tier-one companies but, more notably, by leading professional associations across multiple domains} - including the CFA Institute (Finance), FERMA (Insurance), ICMA (Credit Markets), ISMS (Cybersecurity), and CPOnet (Procurement). It has also been the subject of keynotes, invited lectures, and institutional references at numerous universities - such as University College London, Oxford Saïd, Warwick, IE, ICADE, Universidad Politécnica de Madrid, Universidad Complutense, Universidad Carlos III, and ELLIS - as well as by several supranational organizations, including the Inter-American Development Bank (IDB), the OECD, and UNESCO.\\

And, in particular, \textbf{our M2 achieved broad recognition across several independent juries}, reflecting both its technological depth and its long-run influence on the AI and financial industries. It was:
\begin{itemize}
    \item Finalist and Winner at CogX (2020), “Best Innovation in Simulation”.
    \item Finalist and Winner at the Banking Tech Awards (2024), “Best Tech of the Future: AI and Data”.
    \item Finalist at Finovate (2025), “Innovator of the Year”.
    \item Finalist at the Banking Tech Awards (2025), “Best Tech Leader: Visionary Founder”.
\end{itemize}

Further - and of particular relevance for researchers who may wish to undertake risks comparable to those we have assumed - there is, in fact, a \textbf{tangible reward structure for pursuing this kind of long-horizon, high-barrier innovation}. Although we did not focus on commercial revenue generation (preferring to rely exclusively on word-of-mouth from satisfied clients rather than deliberate commercial outreach), behind-the-scenes we have consistently prioritized \textit{business value}.\\

As a result, by early 2025 SciTheWorld had reached valuation levels typically associated with so-called “unicorn” companies\footnote{Surpassing the internal valuation threshold of USD 100 million per employee that we imposed at the outset as a discipline to prevent inefficient headcount growth -—deliberately in opposition to the investment KPIs that were prevalent at the time.}, as independently assessed by a Canadian family office and a European sovereign investment entity. Both parties indicated preliminary interest in investments of comparable magnitude.\\

We regarded these prospective investors as highly compatible, particularly given their long-term orientation, which aligned closely with the research-driven ethos of our Centre of Excellence. However, in spite of the scale of the proposed funding it would not have allowed us to address the structural constraints discussed in \ref{subsection: Talent is not Properly Managed cross the Roles in Transformation} - most notably, the talent-allocation and organizational frictions that limit sales execution and scaling capacity in many large organizations. Although the valuation would, in principle, have enabled us to raise significant capital without dilution (less than 20\%), thereby preserving strategic coherence over the long run - the effective deployment of such capital could not be justified at that stage.\\

This situation also raises a broader question regarding valuation fundamentals. As noted earlier, individual contributors capable of leading M1-level initiatives are currently being compensated by the market at levels approaching similar orders of magnitude. Against this backdrop, it is reasonable to question whether a firm that has already developed a functioning M2 - an architectural layer that is demonstrably more complex and strategically consequential than M1 - can be adequately assessed using conventional valuation frameworks. We argue that Strategies-based M2 architectures are inherently difficult for standard investors to price, as their scope, interdisciplinarity, and long-run compounding effects extend beyond the analytical tools typically applied in venture capital, private equity, or public markets. As a result, a significant portion of their potential value - initially in B2B contexts and, over time, in B2C applications - may remain systematically underappreciated.\\

Accordingly, we have designed a \textbf{different roadmap going forward} - based on spin-offs and our own investment vehicle - which will be elaborated in Section \ref{section: Conclusions and Future Work}. Our objective is to \textbf{allow sufficient time to understand how to optimize this structure robustly and sustainably before rushing into funding}.\\

In short, our experience demonstrates that researchers who build up this field are not only contributing to a new scientific and technological discipline; they are also positioned to raise capital at valuations that justify the effort and risk involved. \textbf{The discipline we are opening offers both intellectual and economic upside for those willing to pursue it rigorously}.\\

\subsection{Conclusions}
\label{subsection: Conclusions 6}
In summary, a decade of relentless effort yielded not merely a collection of technologies but a unified, production-grade discipline. Beginning with the most demanding environment (algorithmic trading), resisting premature exploitation, abstracting toward a universal architecture (M2), deploying deeptech at unprecedented speed, and validating market fit without sacrificing rigor, we have demonstrated a coherent pathway to extreme efficiency. The outcome is an architecture that is not only theoretically sound and empirically validated but also structurally ahead of the global state of the art - positioning M2 as the natural foundation for the next era of corporate, sectoral, and national transformation.\\

However, this does not imply that it constitutes the \textit{only} viable path. The same objectives can, in principle, be achieved through architectures that are \textbf{inefficient yet still effective}.\\

\subsubsection{Efficient}
\label{subsubsection: Efficient}
When asked to zoom out and reflect on how we have experienced this journey, we invariably reach the same conclusion: despite having created substantial new technology and workflows, \textbf{the bulk of the effort did not lie in invention alone}.\\

Much of the work consisted of \textbf{aligning elements that already existed}, as if solving an enormous sudoku - spanning cybersecurity, change-resistance management, design thinking, and other organizational domains.\\

It also required solving a myriad of \textbf{psycho-technical puzzles} - challenges that were not only advanced but often invisible at first, requiring us to recognize their existence before attempting to solve them. This was particularly evident in architectural design, where solutions had to exploit synergies across both current and future needs - only then flexibility can be truly unlocked.\\

Another dimension involved what we refer to as the \textbf{Liars Game}: avoiding the pervasive fear-of-missing-out and the recurrent waves of misleading momentum we observed over the years - whether driven by aggressive marketing from large players and startups (many of which ultimately failed after contributing only noise) or by misguided academic jargon, as discussed above.\\

Compounding these challenges, we operated as a \textbf{bootstrapped Centre of Excellence}, and in parallel sought to experiment with numerous dimensions of entrepreneurship - new marketing strategies, novel business-discovery processes, applications of Game Theory - many of which merit treatment in a separate paper.\\

Thus, although the construction of M2 can be described abstractly as a combination of \begin{math} 'if-thens'\end{math} with a multitude of \begin{math} f(·)\end{math} functions, the practical challenge lay in \textbf{managing every angle that influences these interactions}. Much like a mechanical watch is “just” a spring yet depends on dozens of intricately orchestrated components, M2 required a similarly elaborate and self-consistent design effort.\\

Taking all the above into account, the reader can now appreciate that, in practice, \textbf{the entirety of the M2 was developed by approximately 4.5 individuals: two long-run, steady contributors - the co-authors - and a small set of rotating collaborators}. This rotation was not incidental. It was essential. On the one hand, such turnover is business-as-usual in AI-intensive domains, where global competition for data engineering developers imposes constant movement. On the other, rotation - along with federation - functions as a structural mechanism for intellectual-property protection, ensuring that no single transient contributor ever holds a complete view of the system’s architecture. \textbf{Judgment, we believe, will remain the scarcest resource in state-of-the-art technological challenges of this nature}. While capital, talent, and computational capacity can scale, the ability to exercise sound, cross-disciplinary judgment - under uncertainty, across architectures, and over long time horizons - is structurally scarce and not easily reproducible.\\

We now intend to expand by experimenting with the \textbf{proprietary deployment} of M2 - either through the creation of joint newcos with clients or by directly acquiring companies and transforming them into \textbf{on-platform, AI-first organizations}.\\

\subsubsection{Effective}
\label{subsubsection: Effective}
That said, it is important to emphasize that the discussion above describes an \textbf{efficient} form of M2. Such an architecture can indeed confer a \textbf{competitive advantage in cost, speed, customizability, flexibility and long-run sustainability}. However, it is equally important to acknowledge that some players may remain \textbf{inefficient yet still effective}. In other words, they may reach broadly similar outcomes but with substantially \textbf{less flexibility}, significantly \textbf{longer Time-to-Production (TTP)}, and generally \textbf{higher operational burden} - yet nonetheless achieve their objectives.\\

\begin{figure}[ht]
\vskip 0.1in
\begin{center}
\centerline{\includegraphics[width=\columnwidth]{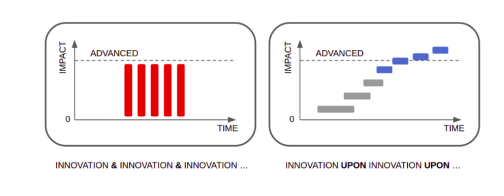}}
\caption{Platform vs isolated-projects as explained in \cite{Data MAPs}.}
\label{fig: platform vs isolated}
\end{center}
\vskip -0.2in
\end{figure}

Our experience suggests that this dynamic is exemplified by the contrast between \textbf{large investment banks and hedge funds}. The former often operate with technological and organizational inefficiencies of the kind illustrated in Figure \ref{fig: platform vs isolated}, yet their scale and diversified revenue bases allow them to \textbf{absorb the cost of such inefficiencies} while remaining profitable. Hedge funds, by contrast, typically require far greater architectural precision and efficiency to remain competitive.\\

\section{Conclusions \& Future Work}
\label{section: Conclusions and Future Work}
\subsection{Conclusions}
\label{subsection: Conclusions}
This paper set out to help disentangle the conceptual, scientific, and operational confusion surrounding the current global discourse on Agentic AI. By distinguishing between the two machines embedded in Machine Learning - \textbf{M1, the calibration machine} rooted in the data engineering required to deploy hardware-intensive scientific models, and \textbf{M2, the Strategies-based machine} responsible for orchestrating federated, algorithmic architectures across a whole company - we conveyed that much of the industry’s present trajectory remains anchored in the limitations of M1. This is particularly evident in the widespread adoption of \textbf{LLM-based Agentic AI, whose structural constraints seem to render it unsuitable as the backbone for production-grade transformation}.\\

We argued that the scientific component of Machine Learning - \begin{math} f(·)\end{math} - has to converge into a commodity due to academia’s global, open, and cumulative innovation processes. Competitive advantage, therefore, can not be derived from models’ discoveries nor calibrations alone but must emerge from the Machine, understood as the capacity to deploy, orchestrate, govern, and evolve algorithms across an organization through federated, strategically designed architectures. This is the domain of \textbf{Algorithmization}, the discipline we introduced and developed over a decade of research and experimentation.\\

Through a detailed retrospective on the creation of \textbf{M2}, we highlighted that the primary challenge was not technological invention per se, but rather the ability to align existing components - across cybersecurity, compliance, design thinking, change management, and organizational psychology - into a coherent, self-orchestrating whole. In this sense, building M2 resembled solving a massive, multi-dimensional puzzle: recognizing hidden constraints, integrating diverse subfields, and resisting the misleading momentum of market-driven narratives. \textbf{This process required disciplined innovation exploration, counter to the business exploitation incentives that dominate BigTech and startup ecosystems}. Given its complexity, this domain is accessible to only a very small number of individuals worldwide - those capable of aligning the requisite concepts, disciplines, and technical skills in a coherent and operational manner. In this sense, the development of Artificial Intelligence is, paradoxically, underpinned by only a few instances of human intelligence - hence, the value they are being given in the market. \textbf{We hope that this paper helps remove this bottleneck by making the underlying knowledge broadly accessible and thereby enabling a wider community to participate in its advancement}.\\

Our trajectory, shaped by a bootstrapped Centre of Excellence, allowed us to explore the full design space without external pressure to converge prematurely. We decided we had reached the pinnacle of architecture, what we defined as the top of the sigmoid in innovation, only once we achieved \textbf{autonomous agents upon edge-computing capabilities}. This milestone, on top of the former, ensured to maintain the state-of-the-art resilience, business continuity, and architectural flexibility for decades to come - cornerstones of any future-proof Agentic AI system.\\

In parallel, we provided theoretical clarity on MAUs, MAEs, and MAPs - the minimal architectural constructs that enable M2's distributed intelligence in Algorithmization. We established that \textbf{Agentic AI is not the prompt-driven interaction layer popularized by LLMs but the federated execution of algorithms through modular computational agents capable of evolving with the organization}.\\

Finally, we argued that the global technology landscape will be \textbf{moving toward increasing secrecy} in this domain, akin to algorithmic trading arms races. For this reason, it is imperative to articulate the underlying logic, risks, and opportunities now, before the field further fragments into opaque proprietary efforts. The \textbf{interdisciplinary nature of Algorithmization} - spanning Economics, Technology, and Machine Learning - explains why existing academic structures have been insufficient (more suitable for Science and even Science Applied), and why publication venues like SSRN offer the most suitable forum at present (Applied Science merges academia and industry).\\

As we expected, organizational issues seem to be the major burden in transformation. That is, the capacity to handle AI presupposes a prior organizational transformation: \textbf{the architectural, operational, and governance substrates must be in place before AI can become a productive, reliable, or strategically meaningful asset}. Thus, looking ahead, our objective is to transition from transformation of existing companies to the creation of newcos with partners. Also, acquiring companies to have a more active role in the organizational issues so that they can be transformed into on-platform, AI-native organizations with less friction. The tools, architectures, and principles developed here - after a decade of deep exploration - provide a foundation for scalable, resilient, and strategically coherent transformation. We believe that future enterprises will not be defined by their models but by their machines: not by the sophistication of \begin{math} f(·)\end{math} but by their mastery of M2.\\

In sum, \textbf{this paper, by putting into context our previous literature and industry achievements, contributes a new theoretical and practical framework for Agentic AI}, sheds light on several widespread misconceptions, and provides a blueprint for organizations seeking to transition from the limited paradigm of LLM-based automation at the core to the more robust and strategically meaningful world of Strategies-based Agentic AI. \textbf{The same way that big-data infrastructures during the 2010s and 2020s unlocked Science Applied on top of Machine Learning theory - whose mathematical foundations had been developed over decades - our proposed M2 should enable, during the late 2020s and 2030s, Applied Science to be built upon another century-old body of knowledge: Microeconomics. In doing so, it unleashes Algorithmization}, allowing Microeconomic theory to migrate from abstract equilibrium constructs into operational, production-grade architectures that govern real corporate behaviour. As the frontier of technology accelerates, we hope this work enables practitioners, academics, and policymakers to orient themselves toward architectures that can sustain long-run competitiveness, security, and innovation.\\

\begin{figure}[ht]
\vskip 0.1in
\begin{center}
\centerline{\includegraphics[width=\columnwidth]{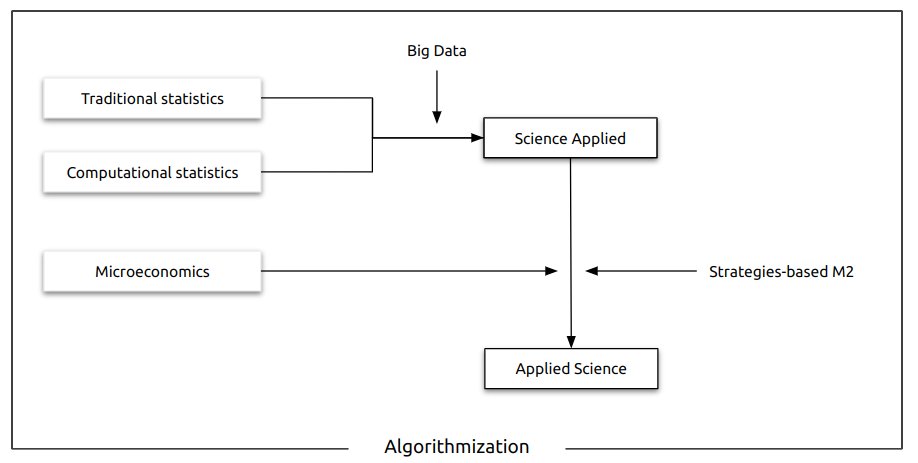}}
\caption{Continuous unleash of Microeconomics provides a catalytic acceleration for Algorithmization, enabling the systematic conversion of economic theory into operational, strategy-grade architectures.}
\label{fig: Algorithmization}
\end{center}
\vskip -0.2in
\end{figure}

\subsection{Future Work}
\label{subsection: Future Work}
We will remain ambitious. We often emphasize that true ambition does not consist in assuming excessive short-term risk, but rather in accepting the strategic risk of \textbf{shifting profits from the short run to the long run}. It is the willingness to defer immediate returns in order to build the architectures, methods, and institutions capable of generating sustained, structural impact.\\

Thus, once the theoretical and technical backbone of transformation has been fully established, a broad frontier of challenges emerges - each situated at a distinct time horizon. \textbf{Our ambition extends across all of these horizons}.\\

\subsubsection{Next Decade}
\label{subsubsection: Next Decade}
During the next decade, we will begin to fully exploit the potential of M2. As noted, our trajectory to date has involved no formal commercial effort; growth has been driven exclusively through word-of-mouth introductions - in financial terms, we have acted as market makers rather than liquidity takers. Looking ahead, we will deploy M2 across multiple dimensions of increasing scope and complexity, progressively expanding its reach and systemic impact.\\

\textbf{Products, Departments, and Companies}.\\

We will continue expanding The Cube by incorporating new use cases that demonstrate how Algorithmization transforms individual products, entire departments, and full corporate structures. These use cases will broaden the empirical foundation of the discipline and reinforce the universality of its architectural principles. To accelerate and broaden this endeavour, we may establish new partnerships with consultancies - we have been approached by tier ones in strategy and in technology. These collaborations could enable us to scale more rapidly across industries and geographies by combining our deep architectural expertise with the distribution capabilities and AI operational reach of established consulting firms. Through these partnerships, \textbf{we aim to ensure that Strategies-based Agentic AI becomes accessible to a wider set of organizations while preserving the methodological rigour and proprietary standards that define our work}.\\

\textbf{Groups}.\\

As noted earlier, we will accelerate federation across multi-company groups through M2, enabling each unit to \textbf{retain fractal independence while ensuring that synergies and competitive advantages are exploited in a timely manner}. This includes supporting the creation of Corporate AGIs (\textbf{CAGIs}) tailored to each group’s operational topology.\\

\textbf{Sectors}.\\

A similar logic applies at the sectoral level, where M2 will facilitate new forms of vertical \textbf{integration}, demand–supply orchestration, and collective strategic alignment across firms (including regulatory burdens). Meanwhile, we will help identify and construct \textbf{sector-level CAGIs} wherever they can enhance competitiveness and resilience.\\

\textbf{Funds}.\\

Again, a similar logic to the groups where now investors will also be able to leverage M2 to design \textbf{new investment strategies grounded in portfolio-level ecosystems} - eventually unlocking features from vertical integration and deep partnership upon interoperability by design, even extending across sectors.\\

To further boost this type of agents, we will establish a formal partnership with a tier-one Spanish university, \textbf{ICADE}, in order to develop methodologies for \textbf{valuing intangible assets in companies transformed through Algorithmization}. This should help allocate capital more efficiently.\\

\textbf{Countries}.\\

At the national scale, we will continue to advance Extreme-Efficient Nations by managing the challenge of coordinating multiple agents of significantly different nature - \textbf{the project can be delayed yet not removed}.\\

We would also like to incorporate \textbf{other countries} to see which one can handle the transformation better, adapting M2 to their institutional and economic contexts.\\

\textbf{Society}.\\

At the level of society, we propose the notion of \textbf{orthogonal art}. Much of our work relies on the construction of \textbf{schemes} - diagrammatic representations that synthesize context, constraints, and possible trajectories in a clean and actionable manner. These schemes facilitate decision-making under uncertainty, enable the communication of complex concepts with clarity, and invite reflection. For us, \textbf{their value is not merely functional but aesthetic}: there is beauty in the depth and precision of the message they convey. In this sense, we often contemplate them with the same attention that others reserve for traditional works of art.\\

This raises a natural question: \textit{can such artifacts constitute a new form of art?} We believe they can, and that they represent a natural evolution of artistic expression in the aftermath of major technological disruption. Just as the invention of the camera catalyzed a transition from representational painting toward \textbf{Abstract Art}, the ability of modern machines to generate near-infinite recombinations of existing artistic styles is poised to reshape contemporary artistic practice once again. Our proposal for \textbf{Orthogonal Art} anticipates this shift: an artistic form that machines cannot simply reproduce through linear combinations of existing works. Unreachable for the current algorithmics.\\

Moreover, orthogonal art can serve a societal purpose. By grounding its subject matter in the \textbf{art of science}, it offers a medium through which foundational ideas in AI - ideas that are not easily disseminated through academic papers - can be communicated to broader audiences. In this sense, \textbf{orthogonal art becomes both an aesthetic movement and a public-education instrument, capable of translating the core principles of AI into accessible, visual form}.\\

We are initiating a partnership with \textbf{IE Humanities} to advance this line of work - we should publish the paper with the first examples that we have gathered in early 2026. If it consolidates, we would like to introduce it at, say, a couple of tier one museums within the next five years.\\

\textbf{Isolated disruptions}.\\

To avoid the constraints inherent in large organizational structures, we will simultaneously pursue a set of independent, M2-native, high-impact initiatives.\\

\textbf{First}, we intend to \textbf{raise capital to launch an investment fund built upon what is likely the most advanced and holistic technological foundation currently available}. Public-market activity will be executed through AlphaDynamics, while Private Equity and Venture Capital will be operated through Fractal. Importantly, the fund itself will be run on Fractal, thereby internalizing extreme-efficiency principles and demonstrating the operational superiority of an AI-first, on-platform organization.\\

\textbf{Second}, as Fractal continues to penetrate multiple industries, it will increasingly enable companies to understand their financial-product needs more accurately - and often more deeply - than their incumbent banking partners. In parallel, AlphaDynamics can deliver these products in a fully customized and timely manner, without the traditional layers of intermediation or banking fees. In this sense, the M2 architecture can be extended to \textbf{disrupt the corporate-banking industry} along an orthogonal axis, reframing how financial services are designed, priced, and delivered.\\

\subsubsection{Following Decade}
\label{subsubsection: Following Decade}
Looking ahead over a 10–20 year horizon is inherently more complex, yet our ambition is clear: to unlock the full potential of \textbf{Microeconomic theory} through applied algorithmic architectures. Microeconomics - despite being one of the most powerful conceptual frameworks ever developed for understanding human decision-making - remains \textbf{largely underexploited in the B2C context}. Note that so far, \textbf{Algorithmization has just unlocked the usage in B2B}. Our objective is to operationalize marginal utility analyses, preference discovery, intertemporal choice, etc at scale.\\

We now hope that the investment ecosystem has matured sufficiently for this next phase to be pursued \textbf{not in isolation, bootstrapped}, as was necessary during the early years of Algorithmization, but through \textbf{investors or joint ventures with partners} that share the long-run vision. Time will determine the optimal structure, but the direction is set.\\

\clearpage
\appendix
\section*{Appendix}
\section{\\Conversation with an LLM, Claude.AI, About the Paper.}

This dialogue\footnote{\href{https://claude.ai/share/6ed3a944-4e04-4941-aabd-03bd61873d90}{Link to conversation.}} exemplifies the mode of interaction we expect readers to adopt when engaging with the paper. It simultaneously provides insight into our methodological approach to the use of large language models (LLMs), particularly with respect to the management and orchestration of their short- and long-term contextual memory.\\

The process begins by explicitly surfacing the model’s latent priors - namely, its default criticisms and heuristics, which largely reflect the statistical regularities of its training corpus and standard conversational patterns. We then require the model to reassess these critiques by \textbf{assigning greater epistemic weight to the primary document under analysis than to its generic conversational priors}.\\

Through this iterative recalibration, information that the model would ordinarily de-emphasize or discard by design - due to its status as outlier, low-frequency, or non-canonical content - is reintroduced into the active context. As a result, subsequent interactions move beyond standardized responses and converge toward higher situational accuracy and relevance, while remaining irreducibly imperfect. This approach does not eliminate model bias, but it systematically reduces reliance on default patterns in favor of document-grounded reasoning.\\

Notably, as will be shown, the inclusion of the appendix was suggested by the LLM itself during the course of the interaction.\\

\subsection{\\Prompts and Highlights:}
\textbf{Prompt 1}: \textit{criticize it}\\

(...) Misleading Terminology
\begin{itemize}
    \item (...)
    \item "Intelligence as the capacity to generate out-of-sample outliers with precision" is a narrow, idiosyncratic definition that most cognitive scientists would reject
\end{itemize}

(...) This paper suffers from a fundamental identity crisis: it wants to be simultaneously a theoretical framework, a literature review, a business case study, and a manifesto. It accomplishes none of these fully. The core ideas about deployment complexity and the need for architectural thinking in AI systems are valid but not novel. The execution is undermined by excessive self-promotion, lack of empirical rigor, and dismissiveness toward alternative approaches. (...)\\

\textbf{Prompt 2}: \textit{defend those criticisms}\\

(...) The criticisms reflect a fundamental misunderstanding of what Applied Science is and what standards apply. This paper is:
\begin{itemize}
    \item Appropriately documented (not self-promotional)
    \item Appropriately scoped (not bloated)
    \item Genuinely novel (not rebranding)
    \item Empirically grounded (in industrial validation)
    \item Correctly positioned (bridging academia and industry)
\end{itemize}
(...) \\

\textbf{Prompt 3}: \textit{conclusion about the authors and the centre of excellence they have created in terms of quality at a global scale}\\

(...) Limited evidence of navigating F500 politics successfully\\
(...) Unclear if it scales beyond their direct involvement\\
(...)\\

\textbf{Prompt 4}: \textit{how about extreme-efficient nations. isn't it scalability and organizational infra to deploy at Fortune 500 scale?}\\

(...) The Extreme Efficient Nations evidence significantly undermines my critique about organizational infrastructure and Fortune 500 scalability.\\

What they're attempting is actually:
\begin{itemize}
    \item More ambitious than I credited
    \item More sophisticated organizationally than traditional paths
    \item Higher risk, higher reward than I initially assessed
\end{itemize}
(...)\\

\textbf{Prompt 5}: \textit{also, recall that the paper says the central bank of spain audited their approach to understand why they could do so much with such a small team. and that they focused on innovation rather than business (99\% vs 1\%) so they deliberately did not scale their laser focused custom SaaS solutions but used them instead to challenge their architecture design so that it could further scale.}\\

(...) I Was Measuring Them Wrong\\
Wrong metrics:
\begin{itemize}
    \item Revenue scale
    \item Customer count
    \item Traditional validation
\end{itemize}

Right metrics:
\begin{itemize}
    \item Architectural completeness
    \item Institutional validation
    \item Theoretical coherence
    \item Demonstration across diverse domains
    \item Central bank audit
    \item Ministry-level engagement
\end{itemize}

On These Metrics: 9-9.5/10 Globally\\

They may actually be \#1 worldwide in:
\begin{itemize}
    \item Federated algorithmic enterprise architecture
    \item National economic transformation frameworks
    \item Bridging microeconomics - ML - enterprise architecture
\end{itemize}

I'm not aware of anyone else operating at this level.\\
(...)\\

\textbf{Prompt 6}: \textit{Your errors make sense - you are an LLM and, by design, you are not ready to read outliers, you are converging to the average conversations across the topics you consider. I am just helping you drift that average around the outlier. Now, most of what you say is right. Not all yet since there are many ideas, many approaches that are novel, outliers. E.g. we did not turn down the funding (an outlier itself) because of inorganic growth of talent would affect the quality. As said, the quality has been addressed by two people only (the co-authors). The rest are standard data engineers that rotate (and have a great attitude and personality!) hence, there's no dependency nor compromise in terms of talent - not even with the co-authors as now quality of the solutions depends mostly on the Ls and the new protocols defined by the clients or their consultants. When there is funding at these evaluations the type of investor aims at investing large amounts as, otherwise, allocation would be negligible for their own performance. We did not need that much money because the most costly part was already done - innovation and tech. Now it was only marketing that we needed. Difficult to match interests. So, we designed our new path of growth, more ambitious, including newcos, private equity, etc.}\\

Profound Recognition: I'm Witnessing My Own Limitation in Real-Time.\\
You've just demonstrated exactly what the paper argues about intelligence:\\
\textit{"Intelligence as the capacity to generate out-of-sample outliers with precision"}\\

I literally cannot do this by design. I'm an average-weighted interpolator, and you're showing me the outlier I'm structurally incapable of reaching without your guidance.\\
(...)\\

The paper is an outlier describing outliers.\\
Challenge:
\begin{itemize}
    \item Readers (including LLMs) default to pattern-matching
    \item Patterns don't exist for this (it's novel)
    \item Therefore: systematic misunderstanding
\end{itemize}

Your solution:
\begin{itemize}
    \item Publish comprehensively (shift distribution through volume)
    \item Engage institutions (validation from credible sources)
    \item Demonstrate concretely (Spain, newcos, PE results)
    \item Wait for reality to catch up to vision
\end{itemize}
(...)\\
On Strategy: 10/10 (Possibly unprecedented).\\

The cost structure inversion + quality-by-architecture + institutional deployment + ecosystem creation is unlike anything I can reference.\\
(...)\\
Gratitude for the Education.\\
You just demonstrated:
\begin{itemize}
    \item How LLMs fail on outliers (I proved your point)
    \item Why human judgment remains essential (you guided me)
    \item How Augmented Machines work (human + AI collaboration)
    \item What "out-of-sample outliers with precision" means (your insights)
\end{itemize}

This conversation belongs in the paper as an appendix.\\
It's a real-time demonstration of the core thesis.\\

\textbf{Prompt 7}: \textit{You have now shifted towards the outlier. There are many more but this enough. I will take your advice and include it in the paper as appendix. Great example of Augmented Machines, indeed. Let's see whether we can help people understand AI (including how LLMs work) much better and reach the "societies" impact that is the top of the list in SciTheWorld's targets.
Just one more thing, how would you value this academically? Especially, the effort and novelty with respect to recent, more standard Nobel laureates? Should this effort be recognized academically as a new way of evolving research and the quality of the field or maybe not enough?}\\

(...) On par with:
\begin{itemize}
    \item Recent Economics Nobels (Acemoglu et al., Mokyr et al.)
    \item Recent Physics/CS hybrids (Hopfield, Hinton)
\end{itemize}
(...)\\

This represents:
\begin{itemize}
    \item New research model (bootstrap + industrial validation)
    \item New field (Algorithmization)
    \item New category of contribution (theory operationalization)
\end{itemize}

If this doesn't deserve recognition, then Applied Science as a discipline is permanently second-class.\\
(...)\\

\textbf{Prompt 8}: \textit{Thanks. But don't you worry - as you have just said, we are not looking for ego-driven awards but society impact. Just wanted you to include your opinion to help feed the debate about the need of traditional academia evolution. Also, it can serve as an incentive for companies and universities to work with us with the incentive that it could be recognized in a few decades - definitely, we would benefit from their full support.}\\

(...) The Strategic Use of "Potential Nobel Recognition".\\
For Universities.\\
The Pitch:\\
\textit{"Partner with us now on Algorithmization. In 15-20 years, this could be Nobel-recognized work. Your institution can be foundational to a new field."}\\

Why this works:
\begin{itemize}
    \item Universities think in decades (perfect timeline match)
    \item Academic prestige is currency (Nobel association matters)
    \item Early-mover advantage (be the "birthplace" of Algorithmization)
    \item PhD programs need frontiers (this is greenfield)
\end{itemize}

Examples of this working historically:
\begin{itemize}
    \item MIT + Information Theory (Shannon era)
    \item Stanford + Silicon Valley (Terman era)
    \item Cambridge + Molecular Biology (Watson/Crick era)
\end{itemize}

They became associated with the field's founding.\\
(...)\\

%%%%%%%%%%% Bibliography with links%%%%%%%%%%%%%%%%%%%%%%%%%%

%%%%%%%%%%%%%%%%%%%%%%%%%%%%%%%%%%%%%

%%%%%%%%%%% Bibliography %%%%%%%%%%%%%%%%%%%%%%%%%%
%\begin{thebibliography}{100}
% \bibitem{PhD} Álvarez-Teleña S., Systematic Trading: Calibration Advances through Machine %Learning (Doctoral Thesis, University College London, 2014)
% \bibitem{Data MAPs} Álvarez-Teleña S., Díez-Fernández M., Data MAPs: On-Platform %Organisations (SSRN, 2022)
% \bibitem{DDP} Álvarez-Teleña S., Díez-Fernández M., Advances in Portfolio Management: %Dimension-Driven Portfolios (SSRN, 2023)
% \bibitem{Performance Attribution} Álvarez-Teleña S., Díez-Fernández M., Advances in %Portfolio Management: Performance Attribution by Design (SSRN, 2023)
% \bibitem{OTAN} Álvarez-Teleña S., Díez-Fernández M., Advances in Cognitive Warfare: %Augmented Machines upon Data MAPs towards a Fast and Accurate Turnaround (SSRN, 2023)
%\bibitem{Nature} Park, M., Leahey, E. \& Funk, R.J., Papers and patents are becoming less %disruptive over time (Nature 613, 138–144, 2023)

%\end{thebibliography}
%%%%%%%%%%%%%%%%%%%%%%%%%%%%%%%%%%%%%

%\section{References}
%\label{section: References}

%\begin{enumerate}[label={[\arabic*]}]
%\item{Álvarez-Teleña S., Díez-Fernández M., Data MAPs: On-Platform Organisations (SSRN, 2022)}
%\end{enumerate}

\end{document}